\documentclass[aps,prx,superscriptaddress,twocolumn,showpacs,amsmath,floatfix,citeautoscript]{revtex4}
\usepackage[colorlinks=true,urlcolor=blue,citecolor=blue,linkcolor=blue]{hyperref} 

\usepackage{graphicx}
\usepackage{float}
\usepackage{dcolumn}
\usepackage{color}
\usepackage{latexsym,bm}
\usepackage[normalem]{ulem}
\usepackage{multirow}
\usepackage{algorithm}
\usepackage{algpseudocode}

\newcommand{\bra}[1]{\langle #1 |}
\newcommand{\ket}[1]{| #1 \rangle}

\newcommand{\RNum}[1]{\uppercase\expandafter{\romannumeral #1\relax}}

\begin{document}
\hyphenpenalty=5000
\tolerance=1000
\title{Accurate simulation for finite projected entangled pair states in two dimensions}
\author{Wen-Yuan Liu}
 \affiliation{Department of Physics, The Chinese University of Hong Kong, Shatin, New Territories, Hong Kong, China}
\author{Yi-Zhen Huang}
 \affiliation{Wilczek Quantum Center and Key Laboratory of Artificial Structures and Quantum Control, School of Physics and Astronomy, Shanghai Jiao Tong University, Shanghai 200240, China}
 \author{Shou-Shu Gong}
 \affiliation{Department of Physics, Beihang University, Beijing 100191, China}
 \author{Zheng-Cheng Gu}
 \affiliation{Department of Physics, The Chinese University of Hong Kong, Shatin, New Territories, Hong Kong, China}
\begin{abstract}
Based on the scheme of variational Monte Carlo sampling, we develop an accurate and efficient two-dimensional tensor-network algorithm to simulate quantum lattice models. We find that Monte Carlo sampling shows huge advantages in dealing with finite projected entangled pair states, which allows significantly enlarged system size and improves the accuracy of tensor network simulation. We demonstrate our method on the square-lattice antiferromagnetic Heisenberg model up to $32 \times 32$ sites, as well as a highly frustrated $J_1-J_2$ model up to $24\times 24$ sites. The results, including ground state energy and spin correlations, are in excellent agreement with those of the available quantum Monte Carlo or density matrix renormalization group methods. Therefore, our method substantially advances the calculation of 2D tensor networks for finite systems, and potentially opens a new door towards resolving many challenging strongly correlated quantum many-body problems.  
\end{abstract}
\maketitle
\date{\today}
\section{Introduction} 
Tensor network states (TNSs) are fundamentally important %concepts 
in modern physics. They provide us a very powerful and efficient way to encode the low-energy physics of complex quantum systems based on their local entanglement structure~\cite{white1992,vidal2004, verstraete2004_3,verstraete2006_2,peps2004,vidal2007,xie2014,yang2014,fPEPS2010,orus2019}, and have led to great success in condensed matter physics and classical statistic physics \cite{schollw2011,orus2014,jutho2017}, for instance, simulating quantum many-body systems~\cite{white1996,white2003,levin2007,liao2017} and classifying topological phases of quantum matter~\cite{gu2008,gu2009,chen2011,schuch2010,schuch2011}. TNSs have also been extensively applied in many other physical areas including quantum field theories~\cite{verstraete2010,jutho2013,taglia2014,hu2018}, quantum circuits~\cite{shi2008,arad2010,kim2017,miles2019,pan2019}, quantum error corrections~\cite{ferri2014,bravyi2014}, holography duality~\cite{swingle2012,hayden2016}, ab-initial calculations~\cite{bogu2012,naka2013,szalay2015} and others. More recently they have even been expanded to the fields of artificial intelligence such as machine learning~\cite{miles2016,miles2018,han2018,ran2018,cheng2019,gao2019,stokes2019} and language models~\cite{gallego2017, pestun2017}.  
 
Unfortunately, the power of the 2D TNSs, specifically, the projected entangled pair states (PEPS), is %greatly 
seriously hindered due to their complexity, unlike the huge success of one-dimensional matrix product states (MPS)~\cite{white1992,white1993,schollw2011}. The first challenge is the extremely expensive computational cost for accurate simulation, whose scaling is at least as high as $O(D^{10})$ on the square lattices in conventional double-layer contraction schemes~\cite{peps2004,verstraete2008,jordan2008,roman2009,xie2009,xie2012,lubasch2014,lubasch2014_2,vanderstraeten2015,yang2017}, where $D$ is the bond dimension determining the representation ability of PEPS. There are two more difficulties in the practical application of PEPS: general setting of the optimization scheme and evaluation of the expectation values of physical observables such as long-distance correlations~\cite{verstraete2008}. Furthermore, the memory cost is a potential bottleneck because it scales as $O(D^8)$.
 
To overcome these challenges, various concepts and algorithms have been developed~\cite{jordan2008,jiang2008,corboz2010,cluster2011,roman2009,pizorn2011,corboz2011,xie2009,xie2012,lubasch2014,lubasch2014_2,phien2015,vanderstraeten2015,huang2016,zhao2016,corboz2016,vanderstraeten2016,xie2017,yang2017,fishman2018,corboz2018,rader2018,vanderstraeten2019,liao2019,hyatt2019}. 
However, the computational cost for accurate PEPS simulation of 2D systems still severely limits the application of PEPS. In particular, as the MPS-based methods strongly suffer from small system size, a breakthrough in PEPS methodology is urgently desired for understanding 2D correlated systems as well as relevant applications such as in machine learning~\cite{miles2016,miles2018,han2018,ran2018,cheng2019,gao2019,stokes2019}.
Among various optimization algorithms, variational Monte Carlo (VMC) sampling may provide an elegant framework to overcome the above major difficulties. In the VMC scheme, the computational cost is reduced to $O(D^6)$ and memory cost is reduced to $O(D^4)$, because one only needs to deal with single-layer tensor networks. It also allows a gradient-based method for accurate optimization, and all kinds of physical observables can be evaluated by MC sampling with the aid of massive parallelization~\cite{sandvik2007,schuch2008}. The proof-of-principle combination of VMC and TNS has been demonstrated in one and two dimensions~\cite{sandvik2007,schuch2008}. However, in the last decade this scheme has received little attention~\cite{wang2011,liu2017,liu2018,dong2019}. 
Recent studies show  one can deal with systems up to $16\times16$~\cite{ liu2017,liu2018}, but it is very challenging to compute larger systems because of the heavy cost of MC sampling. The potential power of this approach is far from clear.

In this paper, we show that the optimized VMC-PEPS method can be very powerful and provides an excellent solution to the above issues. By exploiting the intrinsic advantages of MC sampling, we dramatically speed up the MC sampling efficiency and naturally incorporate it with spin symmetry, resulting in huge improvements for the tensor network calculations. Based on the spin-1/2 $J_1$-$J_2$ Heisenberg model on square lattices, we first demonstrate our method in two cases: the unfrustrated case up to $32\times 32$, which can be unbiasedly simulated by the quantum Monte Carlo (QMC) method~\cite{sandvik1997,sandvik2002}, and the frustrated case up to $8\times 28$ where QMC fails but the density matrix renormalization group (DMRG) method works well~\cite{white2007}. The obtained results for both frustrated and unfrustrated cases agree excellently with available QMC and DMRG results. Finally, to further demonstrate the power of our method, we simulate the frustrated case up to $24\times 24$, where both DMRG and QMC fail. Therefore, our VMC-PEPS method provides a powerful way to resolve many long-standing hard 2D quantum many-body problems as well as machine learning-related problems~\cite{miles2016}.  

The rest of this paper is organized as follows. In Sec.\RNum{2}, we first review the scheme of VMC-PEPS method and introduce the sequetially flipping approach to generate sampling configurations.  Next we show an example to compute the ground state of Heisenberg model with $32\times 32$ sites by gradient optimization method, and compare the Heisenberg model results with QMC and analyze the finite size scaling with open boundary conditions.  In Sec.\RNum{3}, we apply our algorithm to frustrated models. We first consider the square-lattice frustrated $J_1$-$J_2$ model up to $24\times 24$ sites, and compare the results with DMRG or other available results.  Then we present some examples to demonstrate the generality of our method on the furstrated models on triangular and kgaome lattices. In Sec.\RNum{4}, we dicuss the advantages of finite PEPS algorithms and  potential applications to machine-learning problems.

\section{The algorithm} 

\subsection{Monte Carlo Sampling}
 For a quantum spin model on a square lattice with size $N=L_y \times L_x$, the wave function in the form of PEPS with a bond dimension $D$ is given as follows~\cite{peps2004}:
\begin{equation}
 \setlength{\abovedisplayskip}{3pt}
 \setlength{\belowdisplayskip}{3pt}
  |\Psi\rangle = \sum_{s_{1} \cdots s_{N}=1}^P { \rm Tr} (A_{1}^{s_{1}}A_{2}^{s_{2}}
  \cdots A_{N}^{s_{N}})|s_{1}s_{2} \cdots s_{N} \rangle,
   \label{Eq:PEPS}
\end{equation}
where $A_{k}^{s_{k}}=A_{k}(s_{k},l,r,u,d)$ is a rank-5 tensor  residing on site $k$, with one physical index $s_{k}$ whose degree of freedom is $P$ and four virtual indices $l$,$r$,$u$ and $d$ connecting to four nearest-neighbor sites. The dimensions of the virtual indices are $D$. %The function ``Tr'' represents the summation over all virtual indices. 
Without loss of generality, we assume all elements of $A_k^{s_k}$ are real numbers throughout this paper.

For a given TNS, the physical quantities can be computed by using MC sampling over the spin configurations, which was first introduced in Ref.~[\onlinecite{sandvik2007,schuch2008}], based on MPS or string-bond states. Then this scheme was applied to PEPS\cite{wang2011,liu2017}.  
The total energy reads: 
\begin{equation}
 \setlength{\abovedisplayskip}{3pt}
 \setlength{\belowdisplayskip}{3pt}
  E_{\rm tot}= \frac{\langle \Psi |H|\Psi\rangle}{\langle \Psi | \Psi\rangle}=\frac{1}{Z}\sum_{S}{|\Psi(S)|^2E_{\rm loc}(S)} ~~,
 \end{equation}
where the local energy term $E_{\rm loc}(S)$ is defined as 
\begin{equation}
 \setlength{\abovedisplayskip}{3pt}
 \setlength{\belowdisplayskip}{3pt}
E_{\rm loc}(S)=\sum_{S^{\prime}} \frac{\Psi(S^{\prime})}{\Psi(S)} \langle S^{\prime}|H|S\rangle \,.
\label{eq:Es}
\end{equation}
Here $\Psi(S)$ is the coefficient of the configuration $|S\rangle=|s_{1}s_{2} \cdots s_{N}\rangle$ with the form of 
\begin{equation}
 \setlength{\abovedisplayskip}{3pt}
 \setlength{\belowdisplayskip}{3pt}
\Psi(S)=\langle S|\Psi \rangle= { \rm Tr} (A_{1}^{s_{1}}A_{2}^{s_{2}}\cdots A_{N}^{s_{N}}) ~~,
\label{eq:Ws}
\end{equation}
and $Z$ is the normalization factor with $Z=\sum_{S} |\Psi(S)|^2$.
The energy gradients with respect to tensor elements read
\begin{equation}
 \setlength{\abovedisplayskip}{3pt}
 \setlength{\belowdisplayskip}{3pt}
\frac{\partial E_{\rm tot}}{\partial A_{lrud}^{s_k}}=\langle G_{lrud}^{s_k}(S)E_{\rm loc}(S)\rangle-\langle G_{lrud}^{s_k}(S)\rangle \langle E_{\rm loc}(S)\rangle,
\end{equation}
where $\langle \cdots \rangle$ denotes the MC average. $G_{lrud}^{s_{k}}$ is defined as
\begin{equation}
 \setlength{\abovedisplayskip}{3pt}
 \setlength{\belowdisplayskip}{3pt}
G_{lrud}^{s_{k}}(S)=\frac{1}{\Psi(S)}\frac{\partial \Psi(S)}{\partial A_{lrud}^{s_{k}}}=\frac{1}{\Psi(S)} \Delta_{lrud}^{s_{k}}(S) ~~ ,
\end{equation}
where $\Delta_{lrud}^{s_{k}}(S)$ is the element of
\begin{equation}
 \setlength{\abovedisplayskip}{3pt}
 \setlength{\belowdisplayskip}{3pt}
\Delta^{s_{k}}(S)={\rm Tr} (\cdots A_{k-1}^{s_{k-1}}A_{k+1}^{s_{k+1}}\cdots )\, ,
\end{equation}
which is  a four-index tensor obtained by contracting a single-layer tensor network, which excludes the tensor $A^{s_{k}}_{k}$ located on site $k$  on the fixed configuration $|S \rangle$. Therefore, the energy and its gradients with respect to tensor elements can be evaluated by MC sampling, which can be used to optimize PEPS.

  In the MC sampling, we can enforce $\sum_{k}s^z_{k}=0$ for the spin configurations if the ground state of a given spin-1/2 system lives in  the $S^z_{\rm tot}=0$ sector. Generating configurations plays a crucial role in the efficiency of VMC methods.  A popular way to perform the MC sampling is  randomly  choosing a nearest-neighbor antiparallel spin pair (NNASP)  flipping the chosen spin pair to generate trial configurations~\cite{sandvik2010}. Assuming the trial configuration $\ket{S_\beta}$ is obtained by flipping an NNASP which is randomly chosen from the given configuration $\ket{S_\alpha}$, we can get the number of all NNASPs for  $\ket{S_\alpha}$ and $\ket{S_\beta}$, supposing they are $K_\alpha$ and $K_\beta$, respectively. Because each NNASP of $\ket{S_\alpha}$ will be randomly chosen with the same chance $1/K_\alpha$,  the trial configuration $\ket{S_\beta}$ will be accepted with  Metropolis' probability
 \begin{equation}
 P_{\rm r}={\rm min}\Big[1,\frac{|\Psi(S_\beta)|^2/K_{\beta}}{|\Psi(S_\alpha)|^2/K_{\alpha}}\Big] ~~.
 \label{eq:prob_r}
\end{equation}
The next necessary step is to pick a random number $r$  from a uniform distribution on the interval $[0,1)$ and compare it with $P_{\rm r}$. If $r<P_{\rm r}$, $\ket{S_\beta}$ will be accepted as a new configuration; otherwise  $\ket{S_\beta}$ will be rejected and the original configuration $\ket{S_\alpha}$ will be kept. Normally we define a MC sweep as $N$ such flip attempts where $N$ is system size, so that all spins are flipped on average, which can dramatically reduce autocorrelation lengths~\cite{sandvik2010}. The physical quantities such as energy will be measured after each MC sweep. It is easy to verify that the leading cost for generating MC sampling configurations is  $O(N^2D^4D^2_c)$ in each MC sweep, where $N$ is the system size and $D_c$ is the cutoff bond dimension during the contraction process.

   To accelerate the MC sampling,  here we introduce a fast configuration-generating method with cost scaling as $O(ND^4D^2_c)$. Instead of randomly choosing spin pairs to flip, we successively generate configurations by visiting the spin pairs in the order of lattice bonds. In this case, the trial configuration $\ket{S_b}$ obtained  by flipping an antiparallel spin pair from the current configuration $\ket{S_a}$ is accepted with a Metropolis probability 
  \begin{equation}
   \setlength{\abovedisplayskip}{3pt}
 \setlength{\belowdisplayskip}{3pt}
 P_{\rm s}={\rm min}\Big[1,\frac{|\Psi(S_b)|^2}{|\Psi(S_a)|^2}\Big] ~~.
 \label{eq:prob}
\end{equation}
Similar to the random flipping case, we need pick a random number $r$  from a uniform distribution on the interval $[0,1)$ and compare it with $P_{\rm s}$. If $r<P_{\rm s}$, $\ket{S_b}$ will be accepted as a new configuration; otherwise  $\ket{S_b}$ will be rejected and the original configuration $\ket{S_a}$ will be kept. We can visit all of the bonds sequentially and attempt to flip all encountered antiparallel spin pairs residing on lattice bonds according to the probability Eq.~(\ref{eq:prob}).  Such a procedure can also be defined as an MC sweep, in which all spins are flipped on average. Then we can measure the observables after each MC sweep to reduce the autocorrelation lengths. Compared with the random visit scheme, the cost of an MC sweep for sequential visiting can be reduced to $O(ND^4D^2_c)$ by storing some auxiliary tensors, which dramatically improves the efficiency.  We would like to stress that all kinds of lattice bonds must be swept including both horizontal and vertical bonds, and only sweeping one or the other will not produce correct results. The underlying reason is most likely related to the violation of ergodicity.

\begin{figure}
 \centering
 \includegraphics[width=3.2in]{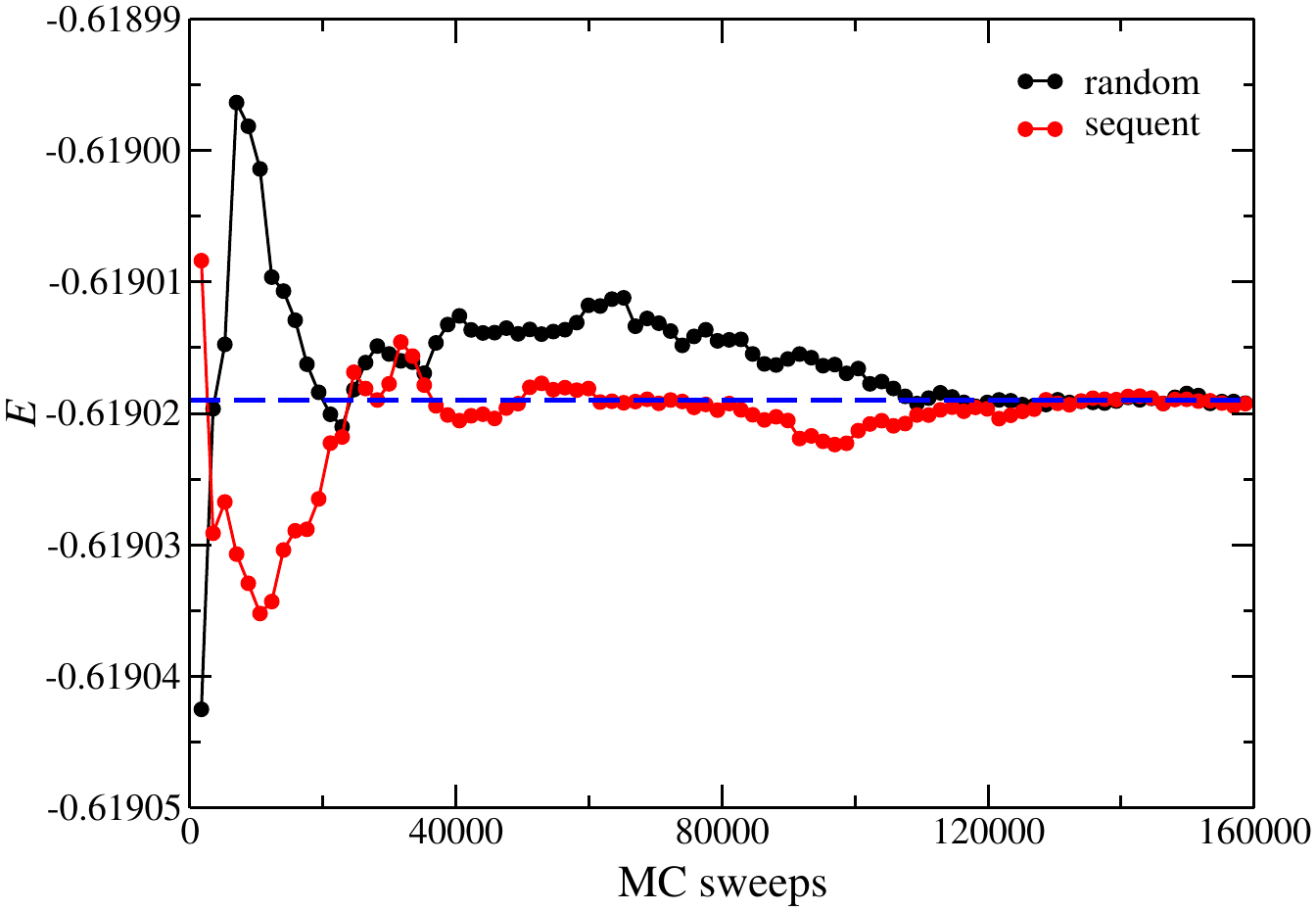}
 \caption{The comparison of energy convergence versus MC sweeps between randomly visiting (black line) and sequentially visiting (red line)  scheme for $8\times 8$ Heisenberg model with $D$=8.  The blue dash line denotes the reference energy $-0.619019$.}
 \label{fig:SweepScheme}
 \end{figure}

 As a direct comparison, we show the energy convergence versus MC sweeps  of the two different visiting schemes for $8\times8$ Heisenberg model with $D=8$, shown in  Fig.~\ref{fig:SweepScheme}. Here we use $D_{c}=24$ to ensure the cutoff convergence for contraction. It shows that in both cases the energies converge very fast, and after about 20000 MC sweeps they converge with errors about $10^{-5}$ even smaller, and the two schemes give the same converged values $-0.619019$ after 120000 MC sweeps. This demonstrates the correctness of the sequentially visiting scheme. 
   
On the other hand, we find a not very large $D_c$ can be good enough to produce correct probability distributions when generating cofigurations, which can further significantly speed up the MC sampling.
Our method involves two steps when evaluating observables such as energies: Generating configurations and computing the local energy $E_{\rm loc}(S)$. The costs of both are dominated by contracting a single-layer tensor network to get $\Psi{(S)}$ scaling as $O(D^4D^2_c)$, and $D_c=3D$ always works very well in practical calculations  for a spin-1/2  $J_1-J_2$ model. We observe that the extremely accurate value of the ratio $R=\frac{|\Psi(S_b)|^2}{|\Psi(S_a)|^2}$ in Eq.~(\ref{eq:prob}) is not very necessary in the process of generating configurations, since it is merely used to compare with a random number. This exhibits the intrinsic advantage of MC sampling. Generally, we always set $D_{c1}=2D$ to generate configurations and $D_{c2}=3D$ to compute $E_{\rm loc}(S)$  for both unfrustrated and frustrated cases. The convergence of $D_{c1}$ and $D_{c2}$ is shown in the Appendices.

 \subsection{Optimization method}
 
Now we turn to the optimization of the PEPS. Optimizing the PEPS wave functions is a very intractable problem. For example, considering a spin-$1/2$  $32\times32$ square system on the open boundary condition (OBC), the number of variables in a $D$=8 PEPS wave function  is as large as about $7.5\times 10^6$, which is a great challenge for optimization. Not only the cost is huge, but also it may be trapped into local minima. To overcome this problem, we first use the simple update (SU) imaginary time evolution method to get a rough ground state for initialization~\cite{jiang2008}, and then use stochastic gradient descent method for further optimization to get an accurate ground state~\cite{sandvik2007,liu2017}.

  In the stochasitic gradient optimization method, each tensor is evolved by a random amount in the opposite direction of the corresponding energy gradient~\cite{sandvik2007,liu2017}, 
\begin{equation}
  A_{lrud}^{s_{k}}(i+1)= A_{lrud}^{s_{k}}(i)-r \cdot \delta(i)\cdot {\rm{sgn}}\Big(\frac{\partial E_{\rm tot}}{\partial A_{lrud}^{s_{k}}}\Big) .
\end{equation}
Here $i$ is the number of evolution step, and $r$ is a random number in the interval $[0,1)$ for each tensor element $A_{lrud}^{s_{k}}$. The parameter $\delta(i)$ is the step length, setting the variation range for an element.

Assuming the optimization step number is $I$ and MC sweep number is $M$, the algorithm structure of the gradient optimization method can be described as follows:
\begin{algorithm}[H]
\caption{VMC-PEPS \label{alg:VMC-PEPS}}
\begin{algorithmic}[1]
\Procedure{VMC-PEPS}{$A$} \Comment{Input tensors $A$}
\While{$i\leq I$} \Comment{$I$ is optimization step number}
	\While{$m\leq M$} \Comment{$M$ is MC sweep number}
		\State generateConfiguration($S_m$) \Comment{Metropolis' algorithm}
		\State  $\Psi(S_m) $=contractNetwork($S_m$)\Comment{ contracting a single-layer tensor network with the configuration $S_m$ }
		\While{ $\langle S^{\prime}_n|H|S_m\rangle \neq 0$} \Comment{sum over possible $S^{\prime}_n$}
				\State  $\Psi(S^{\prime}_n) $=contractNetwork($ S^{\prime}_n$)%\Comment{ Compute $\Psi( S^{\prime}_n)$  by contraction }
		\State $E_{\rm loc}(S_m)=E_{\rm loc}(S_m)+ \frac{\Psi(S^{\prime}_n)}{\Psi(S_m)} \langle S^{\prime}_n|H|S_m\rangle $ \Comment{Compute energy terms}
			\EndWhile
	\State	$G_{lrud}^{s_{k}}(S_m) $=contractDefectNetwork($S_m$, $k$) \Comment{Compute gradient terms}
	\State	$ E_{\rm tot}=E_{\rm tot}+E_{\rm loc}(S_m)$
		
	\State $ P_1=P_1+ G_{lrud}^{s_{k}}(S_m)*E_{\rm loc}(S_m)$
	\State $ P_2=P_2+ G_{lrud}^{s_{k}}(S_m)$
	\EndWhile

    \State $E_{\rm tot}=E_{\rm tot}/M$; $P_1=P_1/M$; $P_2=P_2/M$ %\Comment{average}
 
   \State $P_2=P_2*E_{\rm tot}$;   $ \frac{\partial E_{\rm tot}}{\partial A_{lrud}^{s_k}}=P_1-P_2$ \Comment{ get gradient for site $k$}
   
   \State $A_{lrud}^{s_{k}}(i+1) = A_{lrud}^{s_{k}}(i)-r * \delta(i)* {\rm{sgn}}\Big(\frac{\partial E_{\rm tot}}{\partial A_{lrud}^{s_{k}}}\Big) $ \Comment{update tensors}

	\EndWhile

\EndProcedure
\end{algorithmic}
\end{algorithm}

\begin{figure}
 \centering
 \includegraphics[width=3.2in]{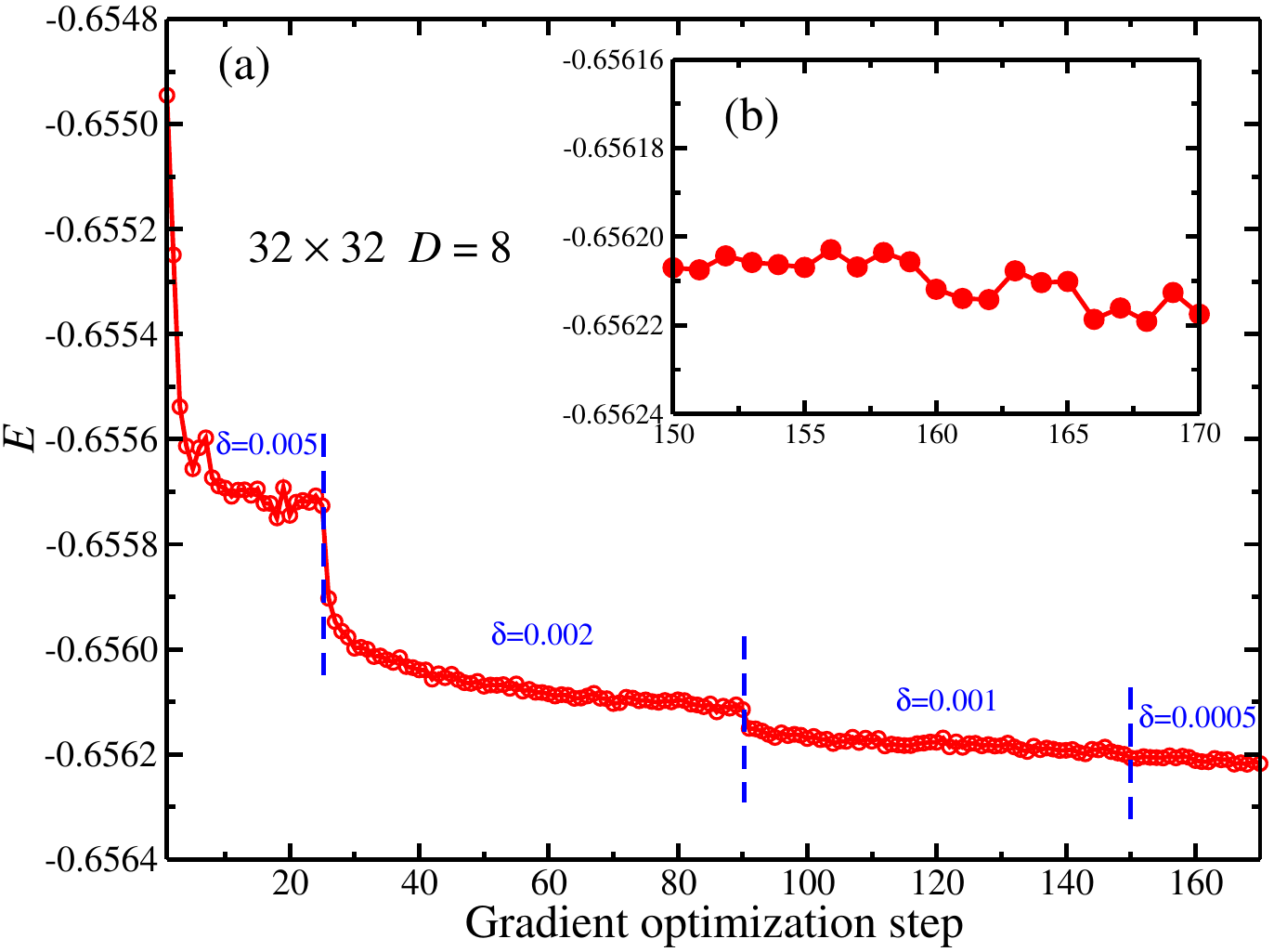}
 \caption{The energy per site varies with the gradient optimization steps for 32$\times$32 AFM Heisenberg model using PEPS $D=8$. (a) In the first 25 steps, set the step length $\delta(i)$=0.005; In the next 65 steps, set  $\delta(i)$=0.002; From the 91-th step to the 150-th step, set $\delta(i)$=0.001; In the last 20 steps, set $\delta(i)$=0.0005. (b) The energy variation in the last 20 steps. }
 \label{fig:GOevolve}
 \end{figure}

Figure \ref{fig:GOevolve} shows the optimization process for a $32\times32$ square AFM Heisenberg model with $D$=8.  We start from the PEPS  obtained by the simple update method. The number of MC sweeps is fixed at 45000 for each gradient optimization step.  In the first 25 steps, the step length $\delta(i)$ is set as 0.005. We can see the energy decreases rapidly at the very beginning, then shows fluctuation. To further decrease the energy, we reduce the step length and keep $\delta(i)$=0.002 until the energy changes slowly.  We can continue to reduce $\delta(i)$ to smaller ones such as 0.001 or 0.0005 for further optimization until the step length has no effects on the energy decrease. Fig.~\ref{fig:GOevolve}(b) shows  the energy variation for the last 20 steps with  $\delta(i)$=0.0005.  The optimization process will be stopped when the energy decreases very slowly. It takes about 4 days with 500 Intel E5-2620 cores for the whole optimization process. 

 Once the optimization process is finished, if one wants to further improve the obtained results, the lattice and spin symmetries  of a given spin-1/2 system can be incorporated into MC sampling~\cite{sandvik2007,sandvik2008}. For example, the spin inversion symmetry (SIS) can be used to get a symmetrized wave function $\ket{\Phi}$ when the optimization is finished  by
${\Phi}(S)={\Psi}(S)+{\Psi}(\bar{S})$
where $S$,  $\bar{S}$ denote the configurations $|S\rangle=|s_1s_2\cdots s_N\rangle$ and  $|\bar{S}\rangle=|-s_1,-s_2,\cdots, -s_N\rangle$, respectively. In general, the results obtained by $\ket{\Phi}$ are slightly more accurate than those of $\ket{\Psi}$. When not otherwise specified, all observables are evaluated with the new weight  $|{\Phi}(S)|^2$ throughout this paper.

\subsection{Heisenberg model on square lattices}

\begin{figure}[htbp]
\setlength{\abovecaptionskip}{0.cm} 
\setlength{\belowcaptionskip}{-0.cm} 
 \centering
 \includegraphics[width=3.2in]{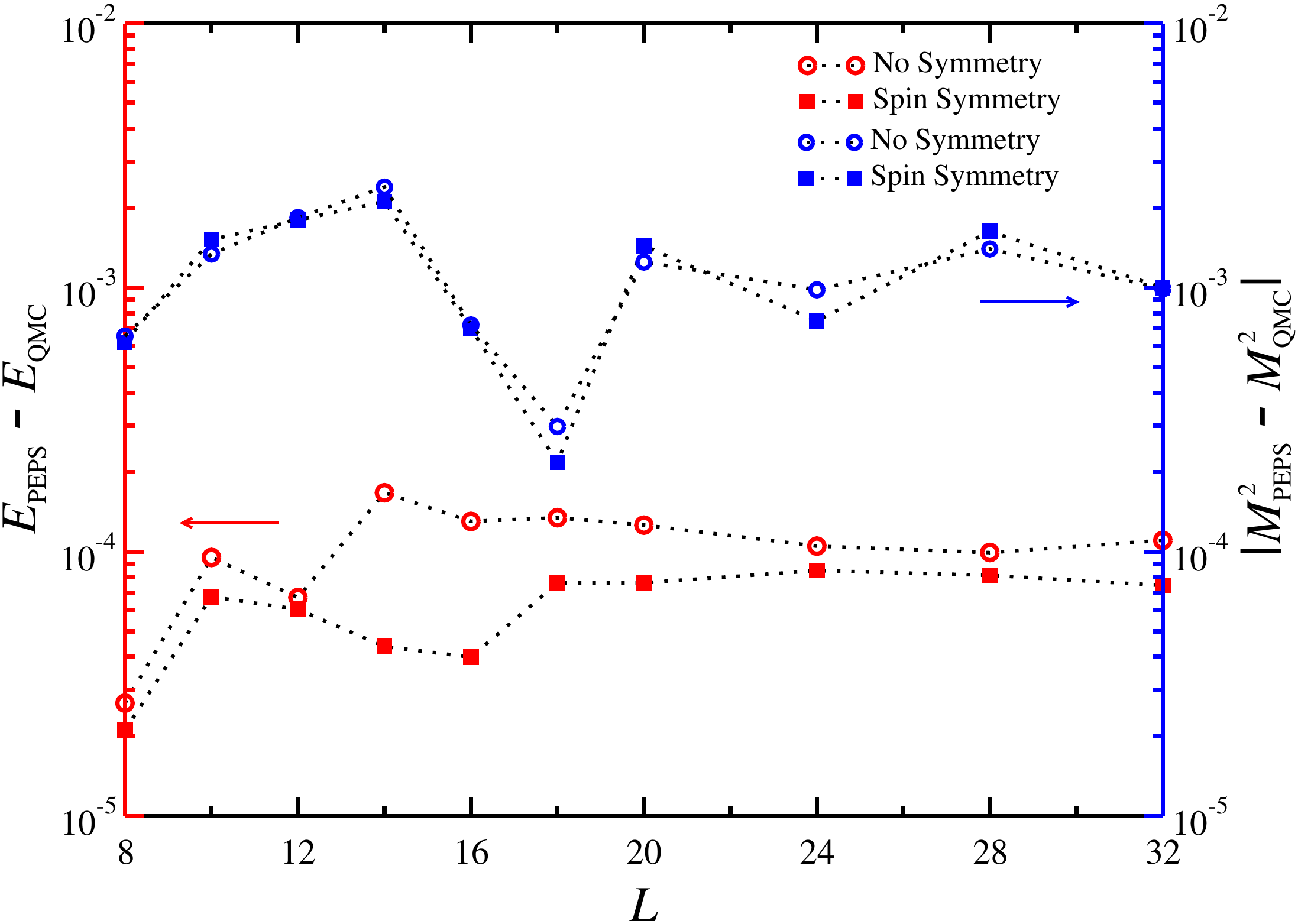}
 \caption{Errors of ground state energy per site and square magnetization for Heisenberg model for $L=8-32$. The red symbols  denote the energy without (open dots) and with (solid squares) spin inversion symmetry. The blue symbols denote corresponding magnetization.  QMC and PEPS  results are listed in Table. \RNum{4}.}
 \label{fig:HeisenbergError}
 \end{figure}
 
% We first consider the antiferromagnetic (AFM) Heisenberg model on an $L\times L$ square lattice, which is unfrustrated and can be unbiasedly simulated by the QMC method~\cite{sandvik1997,sandvik2002}.The ground state energy $E=E_{\rm tot}/N$ and square magnetization $m^2({\bf k})=\frac{1}{N^2}\sum_{ij}{\langle{\bf S}_i \cdot {\bf S}_j}\rangle {e}^{i {\bf k}\cdot({\bf r}_i-{\bf r}_j)}$ are computed with $D$=8 for $L=8-32$. 
The Heisenberg model on an $L\times L$ square lattice provides an excellent test bed for benchmark,  which is unfrustrated and can be unbiasedly simulated by the QMC method~\cite{sandvik1997,sandvik2002}.  We compute the ground state energy $E=E_{\rm tot}/N$ and square magnetization $m^2({\bf k})=\frac{1}{N^2}\sum_{ij}{\langle{\bf S}_i \cdot {\bf S}_j}\rangle {e}^{i {\bf k}\cdot({\bf r}_i-{\bf r}_j)}$ with $D$=8 for $L=8-32$. Fig.~\ref{fig:HeisenbergError} shows the absolute errors of energy and N\'eel AFM  order $M^2=m^2(\pi,\pi)$. We can see that the energies obtained with SIS. are slightly more accurate than those without SIS, but there is almost no difference for the magnetization. When SIS is used, the largest errors of energy and magnetization up to $32\times 32$ are about $8.0\times 10^{-5}$ and $1.0\times 10^{-3}$ those of the QMC results, indicating that our results are highly accurate.  We note that the errors of energy and magnetization vary randomly with respect to $L$, possibly because of incomplete optimization of the PEPS, which contains millions of variational parameters. It takes about 5 days with 500 Intel E5-2620 cores to compute the magnetization for the $32\times32$ lattice with total 50000 MC sweeps.
\begin{figure}[htbp]
\setlength{\abovecaptionskip}{0.cm} 
\setlength{\belowcaptionskip}{-0.cm} 
 \centering
 \includegraphics[width=3.2in]{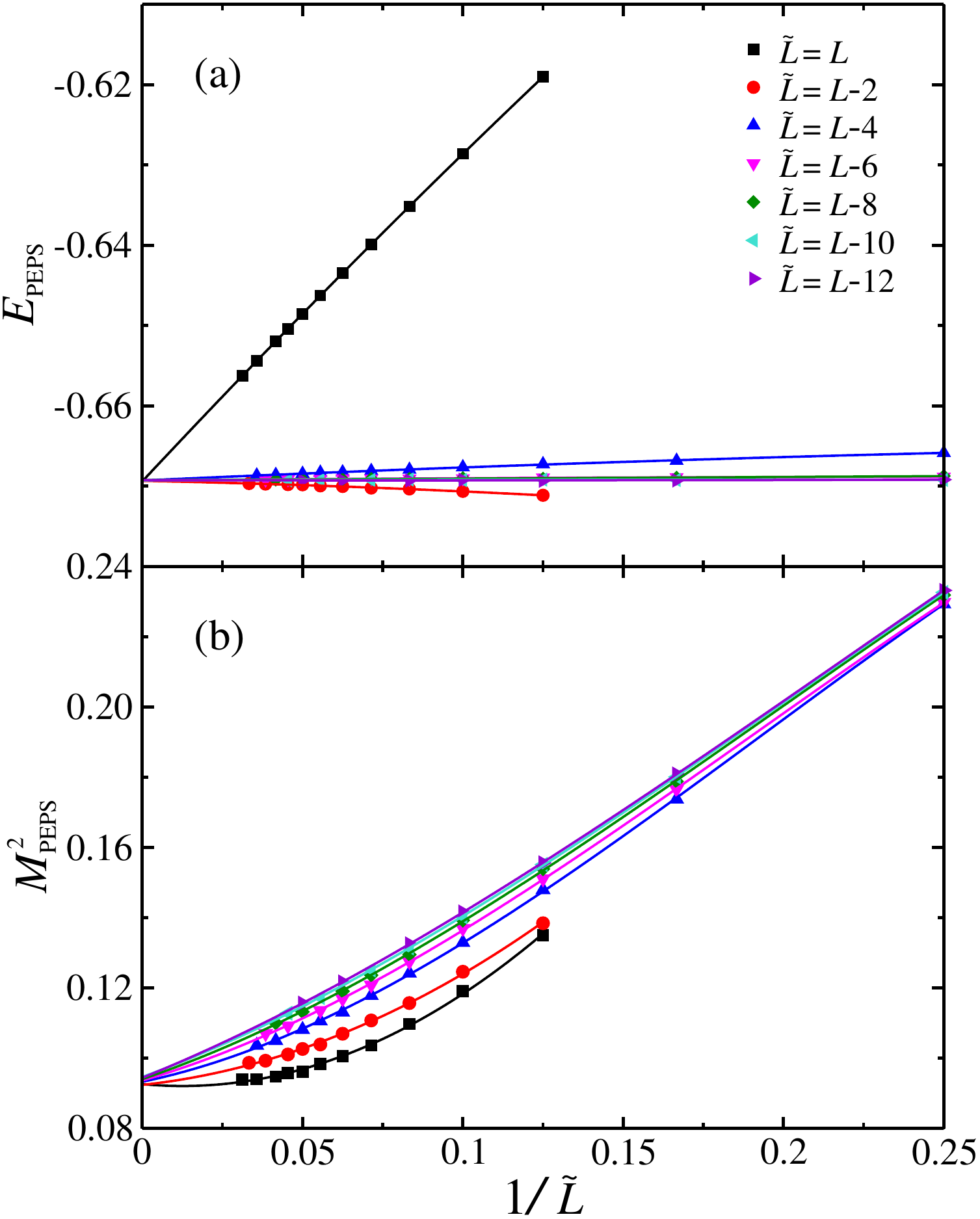}
 \caption{The finite size scaling of ground state energy and magnetization using different choices of central bulk size $\tilde{L}\times \tilde{L}$ on square Heisenberg model with system size $L=8-32$. (a) The extrapolations of ground state energy with different central bulk choices through a second order polynomial fitting. (b)  The extrapolations of magnetization with different central bulk choices. For $\tilde{L}=L$ and $\tilde{L}=L-2$ we use a second order fitting, and for the others a third order fitting is used. Here we use SIS, although one can get similar results without SIS.}
 \label{fig:bulkEnergy_Ms}
 \end{figure}
 
With OBC, boundary effects may play an important role in finite-size scaling (FSS). To reduce boundary effects, one can use the central bulk  $\tilde{L}\times \tilde{L}$ of a given $L\times L$ system~\cite{liu2017}, similar to the standard practice for DMRG calculations when dealing with cylindrical boundary conditions~\cite{white2007}. Various choices are available for the central bulk size $\tilde{L}$ such as $\tilde{L}=L-2$, $\tilde{L}=L-4$ and so on. To systematically investigate the influence of $\tilde{L}$ on FSS, we perform the FSS of energy and magnetization for different $\tilde{L}$ choices, as shown in Fig.~\ref{fig:bulkEnergy_Ms}. We first use the whole system size $\tilde{L}=L$ to perform FSS, in which case the extrapolated energy and staggered magnetization are $E_0=-0.66940(2)$ and $M_0=0.304$. Then we choose the central bulk size as $\tilde{L}=L-2$ and use the corresponding bulk energy and bulk spin order for extrapolation versus the inverse of bulk size $1/\tilde{L}$. Similarly, we can also use other choices such as $\tilde{L}=L-4$. 
The extrapolated values with different bulk choices are listed in Table S1  in Supplemental Material. We can see that all of them excellently agree with  the standard QMC results using periodic boundary conditions (PBC) with $E_{\rm ex}=-0.669437$~\cite{sandvik1997} and $M_{\rm ex}=0.3074$~\cite{sandvik2010_2}.  It is notable that  the boundary effects on energy are significantly reduced when using $\tilde{L}=L-2$. When smaller central bulks are used, we find that both  for the energy and magnetization, the values for different $\tilde{L}$ choices become closer and closer and eventually converge, indicating that boundary effects are gradually eliminated. This is very natural because for a large enough system all bulk choices should give the same value, and specifically for an infinite system there will be no difference among all choices.   Additionally, the bulk energy per site for smaller bulk choices, including $\tilde{L}=L-6$, is approximately $ -0.6692$ and changes very little with respect to system size, providing an efficient way to estimate the energy in the thermodynamic limit.

\begin{table}[htbp]
\caption {Ground state energy $E(\infty)$ and square staggered magnetization $M^2(\infty)$ in the thermodynamic limit using different centall bulk choices $\tilde{L}\times \tilde{L}$ for extrapolations. The numbers in the brackets are  fitting errors. $M(\infty)$ is directly given as the squre root of $M^2(\infty)$ without considering fitting errors. The exact results denote QMC results based on periodic boundary conditions~\cite{sandvik1997,sandvik2010_2}. }
	\begin{tabular*}{\hsize}{@{}@{\extracolsep{\fill}}llll@{}}
\hline\hline
   \multicolumn{1}{c}{\multirow{1}{*}{$\tilde{L}$}}
   &\multicolumn{1}{c} {\multirow{1}{*}{$E{(\infty)}$}}
   &\multicolumn{1}{c} {\multirow{1}{*}{$M^2(\infty)$}} 
   &\multicolumn{1}{l} {\multirow{1}{*}{$M(\infty)$}} \\
   \hline
 		$L$        & -0.66940(2) & 0.0924(11)  & 0.304   \\
 		$L-2$     &-0.66933(6)  & 0.0926(10)  & 0.304\\
		$L-4$     &-0.66929(5)  & 0.0944(06)  & 0.307 \\
		$L-6$     &-0.66928(3)  & 0.0947(08)  & 0.308\\
 		$L-8$     &-0.66926(6)  & 0.0940(13)  & 0.306\\
 		$L-10$   &-0.66928(6)  & 0.0937(13)  & 0.306\\
       	$L-12$   &-0.66927(8)  & 0.0933(12)  & 0.305\\
           exact    &-0.669437~\cite{sandvik1997}  & 0.09451~\cite{sandvik2010_2}        & 0.30743(1)~\cite{sandvik2010_2}    \\
             \hline\hline
	\end{tabular*}
\label{tab:extrapolations}	
\end{table}

\section{Applications to frustrated models}

\subsection{Frustrated model  on square lattices}
\begin{figure}[htbp]
\setlength{\abovecaptionskip}{0.cm} 
\setlength{\belowcaptionskip}{-0.cm} 
 \centering
 \includegraphics[width=3.2in]{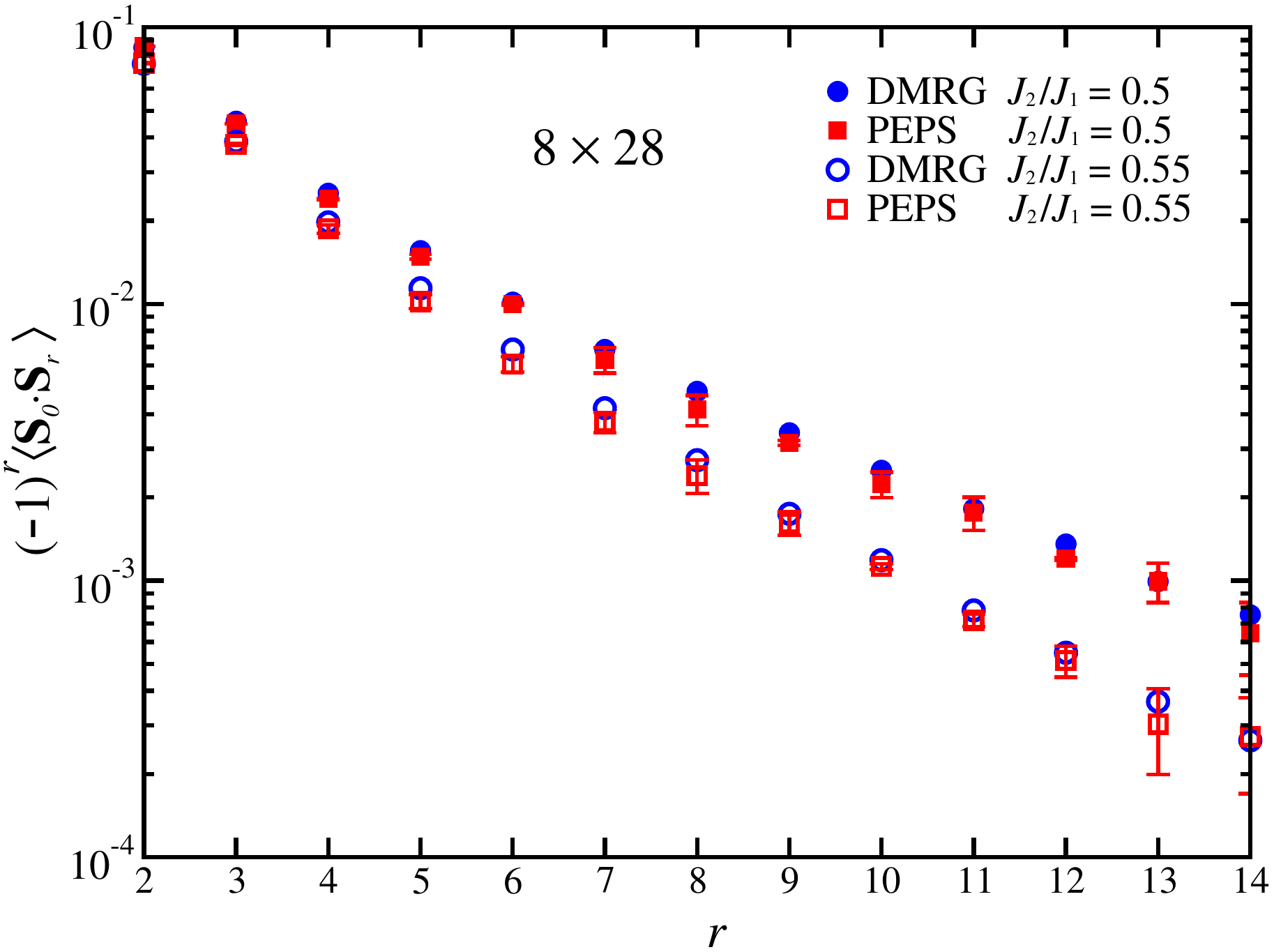}
 \caption{Comparison of PEPS spin correlations with DMRG for $L_y \times L_x$ systems at $J_2/J_1=0.5$ and 0.55 with $L_y$=8 and $L_x$=28. The correlations are measured along the central line $y$=4 and the distance of the reference site away from the left edge is 3 lattice spacings.  }
 \label{fig:spincorrelation}
 \end{figure}

Now we consider the frustrated spin-1/2 $J_1$-$J_2$  square Heisenberg AFM model with size $L_y\times L_x$. In this case, QMC methods suffer the notorious sign problem and we use the DMRG results by keeping 6000 SU(2) states  as references~\cite{gong2014}. We compute an $8\times 28$ stripe at $J_2/J_1=0.5$ and 0.55 with PEPS $D$=8, and the energies  are $-0.488958(1)$ and $-0.478997(1)$, which are very close to the DMRG energies of -0.489036 and -0.479085, respectively; both have an absolute error as small as $8.0\times 10^{-5}$.  Fig.~\ref{fig:spincorrelation} compares the spin correlations with those of DMRG at $J_2/J_1$=0.5 and 0.55. We can see that the PEPS results are in excellent agreement with DMRG. This suggests that our method can also work very well for frustrated models. Here we use 3500000 MC sweeps to obtain long-distance correlation functions with an accuracy of order of $10^{-4}$.
 
\begin{figure}[tb]
 \setlength{\abovecaptionskip}{0.cm} 
\setlength{\belowcaptionskip}{-0.cm} 
 \centering
 \includegraphics[width=3.2in]{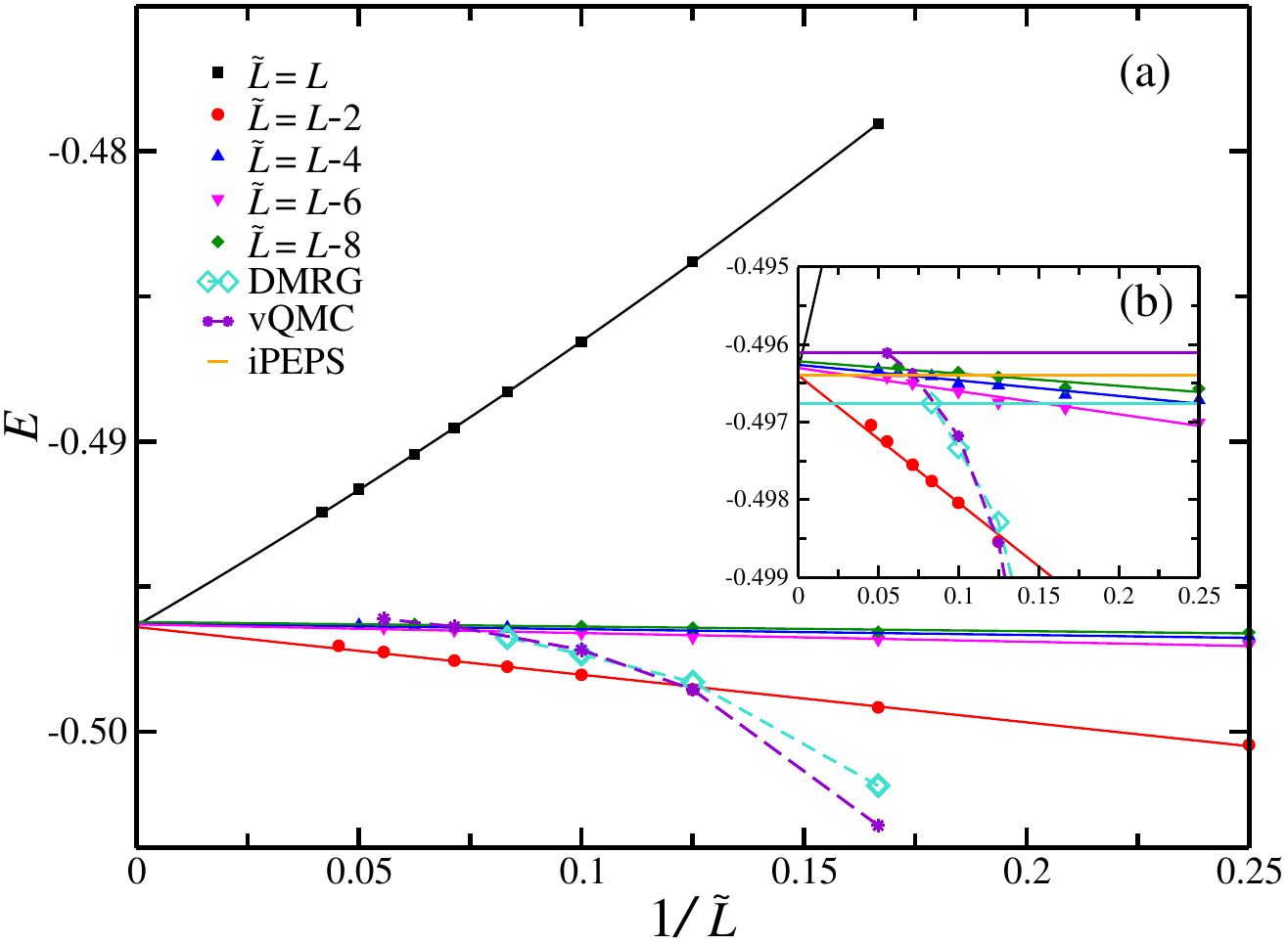}
 \caption{(a) Ground state energies in the thermodynamic limit at  $J_2/J_1$=0.5. Different central bulk choices $\tilde{L} \times \tilde{L}$ are used for extrapolation. The solid lines denote  a second order polynomial fitting for $\tilde{L}=L$ or a linear fitting for the other cases, with extrapolated values lying in a small window [-0.4964, -0.4962]. Here the error bars from MC sampling and fittings are about $10^{-5}$ or smaller. The DMRG  energies on cylinders with circumference $L_c=6-12$ taken from Ref.[\onlinecite{gong2014}], and vQMC energies on torus up to $18\times 18$ with one step of  Lanczos projection from Ref.[\onlinecite{hu2013}], are also shown.  (b) The results on a more detailed scale. The solid cyan line denotes the DMRG energy of $L_c=12$, and  the solid  violet line denotes the vQMC energy of  $18\times18$ sites; both are directly used as extrapolated energies in 2D limit according to Ref.[\onlinecite{gong2014}] and Ref.[\onlinecite{hu2013}], respectively. The orange line denotes iPEPS $D=8$ energy -0.4964 taken from Ref.[\onlinecite{reza2018}].} 
 \label{fig:GSenergyJ05}
 \end{figure}

To further demonstrate the power of our method, we consider large frustrated systems up to $24\times 24$ at $J_2/J_1$=0.5, on which neither QMC nor  DMRG can work. We compare extrapolated ground state energies  in the 2D limit with available results obtained by DMRG~\cite{gong2014} and variational QMC (vQMC) methods~\cite{hu2013}. In Fig.~\ref{fig:GSenergyJ05}(a), different central bulk  $\tilde{L} \times \tilde{L}$ choices are used for extrapolation, and they all produce almost the same extrapolated energies. Specifically, the choice of  $\tilde{L}=L$ gives extrapolated energy -0.49635(5), very close to the extrapolated DMRG energy of -0.4968 and the vQMC plus one Lanczos step energy of -0.4961, as well as an iPEPS $D=8$ energy of -0.4964. The details are shown in Fig.~\ref{fig:GSenergyJ05}(b). 
%We would like to point out that the DMRG energy is obtained from a cylindrical system with 12 lattice spacings as the circumference~\cite{gong2014}, and the vQMC energy is obtained from a periodic $18\times 18$ system~\cite{hu2013},
We would like to point out that the extrapolated DMRG energy and  vQMC energy are both taken from finite systems, so they should be regarded as the corresponding lower bounds of the ground state energies in the 2D limit. In addition, comparing with the energies of different sizes from DMRG and vQMC, bulk energies from OBC show much smaller finite size effects. The above results not only verify that our method can work well on large frustrated systems, but also demonstrate the validity of FSS for both unfrustrated and frustrated systems with OBC. It is notable that previous studies  can deal with systems size up to $16\times16$ and obtain close extrapolated energies~\cite{liu2017,liu2018}, but larger sizes calculations up to $24\times24$ are of great importance to confirm the correctness of those based on smaller systems because they have smaller FSS errors. More results for $J_1$-$J_2$ model are reported elsewhere~\cite{liu2020}.
   
   \begin{table*}[htbp]
   \centering
 \caption {The ground state energies and spin orders for 24$\times$24 at $J_2/J_1$=0.55 with different bond dimension $D$. Here $M^2$=$m^2(\pi,\pi)$ and $M^2_{\alpha}$=$m^2_{\alpha}(\pi,\pi)$ for  $\alpha=x,y,z$, and ${\bf k_x}=(\pi,0)$ and ${\bf k_y}=(0,\pi)$. }
	\begin{tabular}{cccccccc}
		\hline\hline
	$D$& $E$ & $M^2$     & $M^2_x $ &  $M^2_{y}$ &  $M^2_z$ &  $m^2(\bf k_x)$& $m^2({\bf k_y})$     \\ \hline
 		4 & -0.479656(3)   &7.46(4)$\times 10^{-3}$      & 2.12(1)$\times 10^{-3}$ & 1.96(1)$\times 10^{-3}$& 3.38(2)$\times 10^{-3}$  &1.07(1)$\times 10^{-3}$  &0.96(1)$\times 10^{-3}$ \\
 		5 & -0.481498(6)   &6.15(3)$\times 10^{-3}$     & 2.09(1)$\times 10^{-3}$& 1.97(1)$\times 10^{-3}$& 2.09(1)$\times 10^{-3}$ &0.99(1)$\times 10^{-3}$ &1.04(1)$\times 10^{-3}$ \\
		6 & -0.481860(1)   &6.18(8)$\times 10^{-3}$     & 2.03(4)$\times 10^{-3}$& 2.04(2)$\times 10^{-3}$& 2.11(2)$\times 10^{-3}$ &0.99(3)$\times 10^{-3}$&1.08(2)$\times 10^{-3}$\\
 		7 & -0.481976(3)   &6.24(7)$\times 10^{-3}$     & 2.06(2)$\times 10^{-3}$ & 2.06(2)$\times 10^{-3}$& 2.12(3)$\times 10^{-3}$  &1.03(3)$\times 10^{-3}$&1.07(2)$\times 10^{-3}$\\
 		8 & -0.481989(1)   &6.26(8)$\times 10^{-3}$     & 2.07(4)$\times 10^{-3}$&  2.09(2)$\times 10^{-3}$ & 2.10(2)$\times 10^{-3}$ &1.02(3)$\times 10^{-3}$&1.04(4)$\times 10^{-3}$ \\
 		\hline\hline
	\end{tabular}
\label{tab:24x24J055}	
\end{table*}

%\subsection{Convergence of bond dimension $D$}

To check the convergence with respect to bond dimension $D$, we calculate the system $24\times 24$ at a highly frustrated point $J_2/J_1=0.55$ with $D$ increasing from 4 to 8. For each $D$, we first use simple update to get an initial state, and then use gradient method for further optimization until the energy varies little.  Once optimization is finished, physical observable will be evaluated via MC sampling.  Spin orders are computed with single component $m^2_{\alpha}({\bf k})=\frac{1}{N^2}\sum_{ij}{\langle{\bf S}^{\alpha}_i \cdot {\bf S}^{\alpha}_j}\rangle {e}^{i {\bf k}\cdot({\bf r}_i-{\bf r}_j)}$ where $\alpha=x,y,z$, as well as the full component one  $m^2({\bf k})$.   In Table.~\ref{tab:24x24J055}, we list the  ground state eneries and  spin orders including $M^2_{\alpha}$=$m^2_{\alpha}(\pi,\pi)$ and $m^2({\bf k})$. With increasing $D$ from 4 to 8, we can see the energy persit $E$ and spin order converge gradually. Notable  $M^2_{\alpha} $ are getting closer to $\frac{1}{3}M^2$, indicating the SU(2) symmetry is gradually recovered. At the same time, we have $m^2(\pi,0)=m^2(0,\pi)$ for large $D$, implying the isotropy of $x-$ and $y-$ axes is also recovered. These results means our method indeed finds the correct ground state subspace.

\subsection{Frustrated models on triangular and kagome lattices}

\begin{figure}
 \centering
 \includegraphics[width=3.2in]{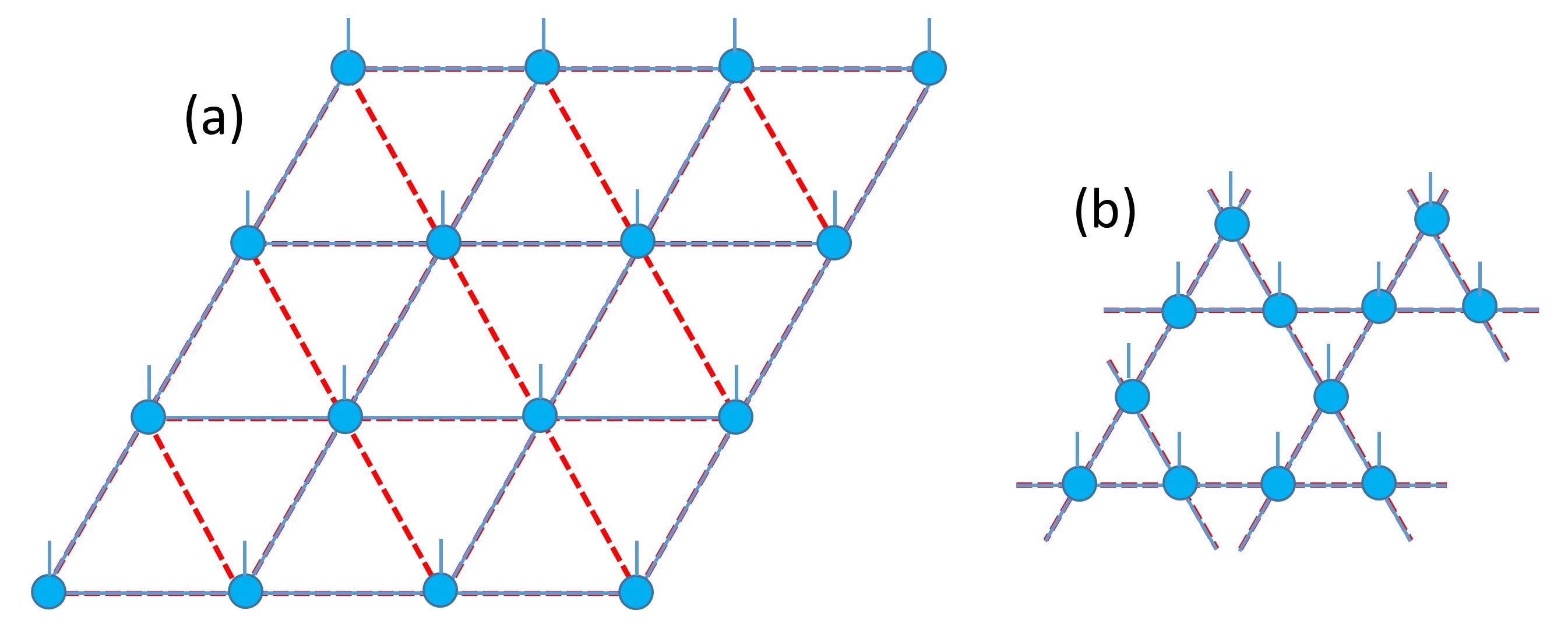}
 \caption{(a) The  square-lattice PEPS representation of  triangular-lattice models. (b) The PEPS representation on a kagome torus with 12 sites.} 
 \label{fig:lattice}
 \end{figure}

To demonstrate the generality of our method on other frustrated models, we present some results on triangular lattice and kagome lattice.
For triangular lattice, we can still define the PEPS  on the square lattice.  Thus one kinds of the next-nearest neighbor interaction terms along a certain direction in the triangular lattice will be seen as nearest neighbor terms for square-lattice PEPS. Then we can directly apply our method  to the triangular models.  We compute the fully open spin-1/2 $J_1$-$J_2$ triangular Heisenberg model on a $4\times 4$ square lattice with $D=8$, shown in Fig.~\ref{fig:lattice}. The results for different $J_2$ are  listed in the left part of Tab.~\ref{tab:lattice} (we set $J_1=1$). We can see all the energies are in excellent agreement with those from exact diagonalization (ED).

For Kag\'ome lattice, here we consider a small torus with 12 sites, and the PEPS are defined on the lattice. Then we can still use the gradient method to optimize the energy function. Since the system is small, we can use exact summation ro replace the MC sampling. We compute the  Kag\'ome  antiferromagnetic Heisenberg model  with nearest neighbor interactions (set coupling constant $J_1=1$). Results of different bond dimension $D$ are listed in the right part of  Tab.~\ref{tab:lattice}. The initialization for different $D$ is always from random tensors.  Except the case of $D=2$, all other cases of $D\geq3$ are the same with ED results. The above results demonstrate that our method can also work well on triangular and Kag\'ome lattices. We leave the large-size calculations for further studies. 
 \begin{table}[htbp]
\caption {The left part is  comparison between PEPS  and ED for the ground state energies of spin-1/2  $J_1$-$J_2$  antiferromagnetic Heisenberg model on triangular lattices. The right part is the ground state energies  of a 12-site Kag\'ome antiferromagnetic Heisenberg model with different bond dimension $D$. }
	\begin{tabular}{ccc||cc}
	%\begin{tabular*}{\hsize}{@{}@{\extracolsep{\fill}}cllll@{}}
\hline\hline
   \multicolumn{1}{c}{\multirow{1}{*}{}}
   &\multicolumn{1}{c} {\multirow{1}{*}{triangular}}
   &\multicolumn{1}{c||} {\multirow{1}{*}{}}  
   &\multicolumn{1}{c} {\multirow{1}{*}{}} 
   &\multicolumn{1}{c} {\multirow{1}{*}{kagome}} \\
    %  \cline{1-3} \cline{4-5} 
   %& triangle  & kagome  \\ \hline
   \hline
    $J_2$ & PEPS &ED & $D$&energy \\
   \hline
 		0     & -0.481851 &-0.4818527 & 2 & -0.4519551985\\
 		0.02 &-0.479451 &-0.4794522 & 3 & -0.4537396014\\
		0.04 &-0.477199 &-0.4772004 & 4 & -0.4537396014\\
		0.06 &-0.475097 &-0.4750979 & 5 & -0.4537396014\\
 		0.08 &-0.473144 &-0.4731453 & 6 & -0.4537396014\\
 		0.10 &-0.471341 &-0.4713428 & 7 & -0.4537396014 \\
       	0.12 &-0.469690 &-0.4696908 & 8 & -0.4537396014\\
           0.14 &-0.468188 &-0.4681894 & ED &-0.4537396014\\
             \hline\hline
	%\end{tabular*}
	\end{tabular}
\label{tab:lattice}	
\end{table}

  \begin{table*}[htbp]
\caption {The  ground state energies and square staggered magnetizations  for Heisenberg model on $L\times L$ square systems for $L=8-32$ on open boundary conditions. QMC results are obtained by the loop stochasitic series expansion (SSE) method with inverse temperature up to $\beta=120$~\cite{sandvik2002}.  $E_w({\rm PEPS})$ and $M^2_w({\rm PEPS})$ denote PEPS $D=8$ results with spin inversion symmetry, and $E_o({\rm PEPS})$ and $M_o({\rm PEPS})$ denote PEPS $D=8$ results without  spin inversion symmetry. }
	%\begin{tabular}{ccccc}
	\begin{tabular*}{\hsize}{@{}@{\extracolsep{\fill}}cllllll@{}}
\hline\hline
   \multicolumn{1}{c}{\multirow{1}{*}{$L$}}
   &\multicolumn{1}{c} {\multirow{1}{*}{$E$(QMC)}}
   &\multicolumn{1}{c} {\multirow{1}{*}{$E_w$(PEPS)}}
      &\multicolumn{1}{c} {\multirow{1}{*}{$E_o$(PEPS)}}
    &\multicolumn{1}{c} {\multirow{1}{*}{$M^2$(QMC)}}
        &\multicolumn{1}{c} {\multirow{1}{*}{$M_w^2$(PEPS)}} 
    &\multicolumn{1}{c} {\multirow{1}{*}{$M_o^2$(PEPS)}}  \\
   \hline

 		8 & -0.619040(1) &-0.619019(1) &-0.619013(1) &0.13565(6) & 0.13503(9)&0.13499(0)   \\
 		10 &-0.628667(2)&-0.628600(2) &-0.628572(4) &0.12062(6) & 0.11910(0)&0.11928(0)\\
		12 &-0.635203(1)&-0.635143(4) &-0.635136(6) &0.1115(2)  &  0.10974(7)&0.10970(9) \\
		14 &-0.639925(3)&-0.639882(3) &-0.639758(4)&0.1058(3)  &  0.10364(8)&0.10337(9) \\
 		16 &-0.643528(8)&-0.643489(1) &-0.643398(9)&0.1014(4)  &  0.10065(9)&0.10062(3) \\
 		18 &-0.646335(2)&-0.646259(2) &-0.646200(0)&0.0985(2)  &  0.09830(5)&0.09822(6) \\
       	20 &-0.648607(9)&-0.648531(5) &-0.648481(0)&0.0976(5)  &  0.09621(4)&0.09644(2) \\
           22 &-0.650463(4)&-0.650387(3) &-0.650336(0)&0.0957(9)    &  0.09578(8)&0.09573(2) \\
           24 &-0.652023(8)&-0.651938(1) &-0.651918(0)&0.0940(8)    &   0.0948(2) &0.0950(0)\\   
           28 &-0.654472(8)&-0.654391(4) &-0.654373(5)&0.092(2)    &  0.0940(1)  &0.0938(1) \\
           32 &-0.65633(1)  & -0.656251(2) &-0.656216(6)&0.095(3)     &  0.0940(2)  &0.0940(0)\\
             \hline\hline
	\end{tabular*}
	
	%\end{tabular}
\label{tab:Heisenberg}	
\end{table*}
 
\section{Conclusion and discussion}

In summary, we developed an accurate and efficient finite PEPS method based on the scheme of VMC. Our method provides a powerful approach to overcome the core difficulties of tensor network simulation encountered in practical applications, and makes it possible to deal with large quantum systems up to $32\times32$ sites with high precision.  For both unfrustrated and frustrated systems, the obtained results are in excellent agreement with QMC and DMRG. In principle, we can handle larger systems and larger bond dimensions by using more computational resources with massive parallelization, which is the built-in advantage of our method. Our method can also be generalized into fermion systems straightforwardly to deal with Hubbard/t-J model by using Grassmann number TNSs. 
 
We would like to stress that the algorithmic development of finite PEPS ansatz is very necessary. Current progress is mainly focused on infinite systems based on the infinite PEPS (iPEPS) ansatz, with a few tensors in a small unit cell, whereas the finite PEPS ansatz allows a different approach without pre-defining a unit cell, providing the complementary part to the  tensor network community. In particular, based on finite 2D systems, it can be directly compared with available DMRG results, which would be very crucial to  clarify the nature of some  controversial quantum many-body problems~\cite{liu2020}. Furthermore, it can also simulate translation-invariance broken systems, including the phases involving short-range or long-range incommensurate orders, as well as trapped cold atoms.  It should even allows us to study the real time evolution in the t-VMC scheme~\cite{carleo2012}.

On the other hand, the finite PEPS algorithm can also benefit  machine learning field. Current tensor network studies on machine learning is mainly focused on one-dimensional MPS and quasi-one-dimensional tree tensor networks, the related algorithms to which have been well established. Intutively, PEPS seems to be a more natural approach  because in physics it has the same geomertic structure as the natural image~\cite{cheng2020}. Due to its complexity, only until last year did a work using PEPS for supervised learning come up~\cite{cheng2020}. However, the optimization of PEPS in that work might be still an issue that affects the accuracy~\cite{cheng2020}. We expect our algorithm would improve the performance of future PEPS application in machine learning models, not limited to supervised learning mentioned above. 

\section{Acknowledgments} We thank F. Yang, D. Poilblanc and A. W. Sandvik for helpful discussion. W.-Y. Liu is also indebted to L. He, Y.-J. Han, S.-J. Dong and C. Wang for correlated work. The method is developed with a Fortran library TNSpackage~\cite{dong2018}. Z.-C. Gu is supported by Direct Grant No. 4053346 and Group Research Scheme(GRS) No. 3110113 from The Chinese University of Hong Kong; funding from Hong Kong's Research Grants Council (NSFC/RGC Joint Research Scheme No. N-CUHK427/18). S.-S. Gong is supported by NSFC grants No. 11874078, 11834014 and the Fundamental Research Funds for the Central Universities. Numerical calculations were carried out on HPC clusters of CUHK and National Supercomputing Centre in Shenzhen. 

\appendix

\section{Simple update}
\begin{figure*}
 \centering
 \includegraphics[width=6.4in]{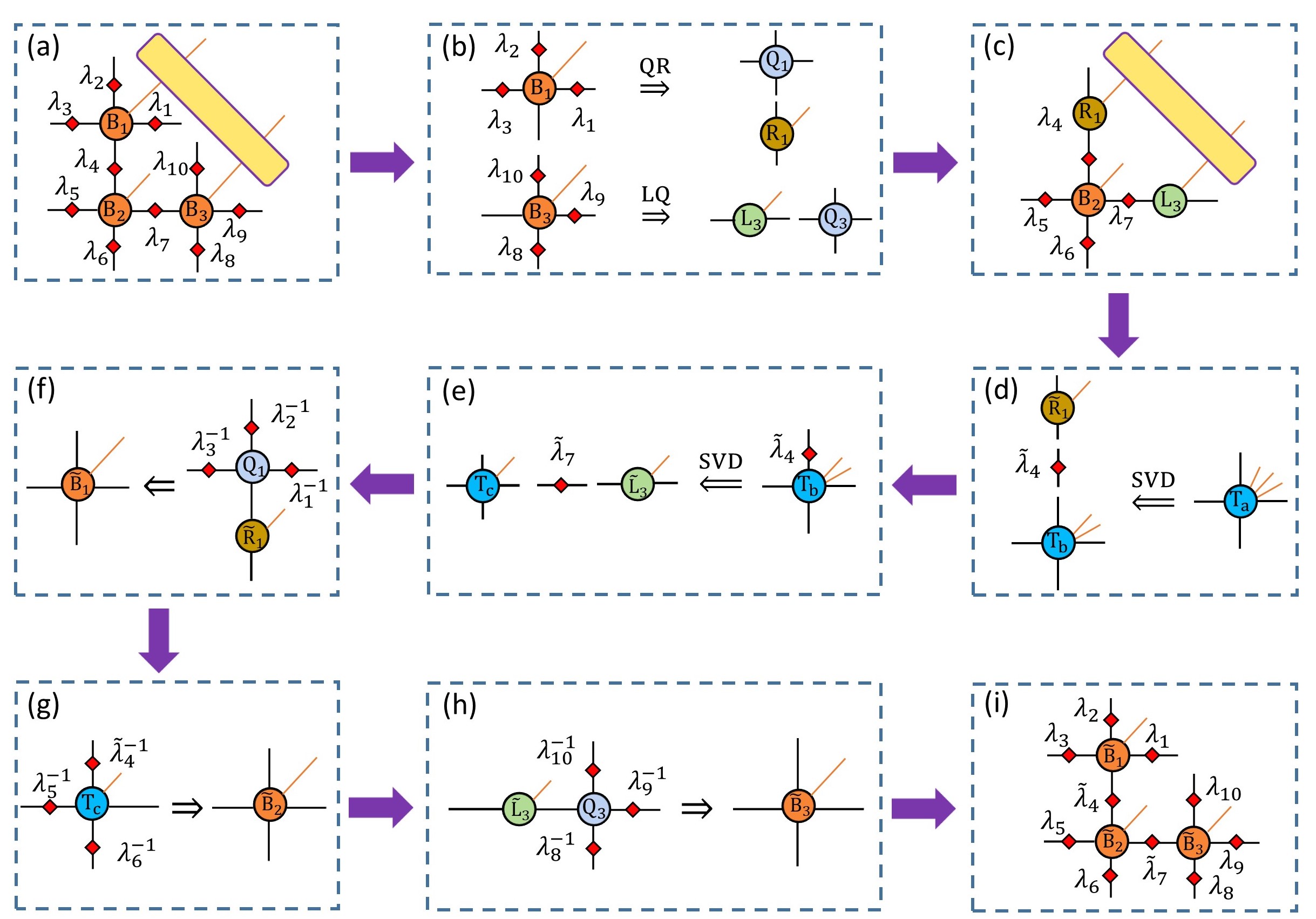}
 \caption{How to update tensors  in the scheme of  the SU method with NNN iteraction. (a)The NNN evolution operator only acts on several local tensors.  In the SU method, PEPS are comprised of tensors $B_i$ living on site $i$ and diagonal matrices $\lambda_i$ living on the link between nearest-neighbor sites . (b) Contract $B_1$ ($B_3$) with  corresponding $\lambda$s, then do QR (LQ) decomposition on the resulting tensor to get $Q_1$  ($Q_3$)and $R_1$ ($L_3$). (c) Contract the NNN evolution operator  with $R_1$, $L_3$, $B_2$ and corresponding $\lambda$s, getting a new tensor $T_a$. (d) Perform SVD on $T_a$ with truncations to get $\tilde {R}_1$, $\tilde{\lambda}_4$ and $T_b$. (e) Contract $T_b$ and $\tilde{\lambda}_4$,  then perform SVD with truncations to get $\tilde {L}_3$, $\tilde{\lambda}_7$ and $T_c$.  (f) Contract $Q_1$ with $\tilde {R}_1$ and the inverse of corresponding $\lambda$s  to get $\tilde{B}_1$.  (g) Contract $T_c$ with the inverse of corresponding $\lambda$s  to get $\tilde{B}_2$. (h) Contract $Q_3$ with $\tilde {L}_3$ and the inverse of corresponding $\lambda$s  to get $\tilde{B}_3$. (i) A new tensor network state with structure invariance is obtained.  }
 \label{fig:simpleupdate}
 \end{figure*}

   In the scheme of simple update, the environment of a local tensor is approximated by a series of diagonal matrices, leading to a cheap cost for updating tensors. When dealing with  the terms of nearest neighbor (NN) and next-nearest neighbor (NNN) interactions, the cost can be reduced to $O(D^5)$ with the help of QR (LQ) decomposition. Fig.~\ref{fig:simpleupdate} shows how to update tensors for NNN interactions. When an NNN evolution operator acts on the given PEPS, what we need to do is updating several involved tensors, shown in Fig.~\ref{fig:simpleupdate} (a).  Fig.~\ref{fig:simpleupdate} (b)-(i) depicts how to update tensors using QR (LQ) decomposition.  For detailed explanations please see the caption of Fig.~\ref{fig:simpleupdate}. 
    
    When using SU method to update tensors, given the bond dimension $D$, we usually start with a imaginary time step $dt=0.01$ until the environmental tensors $\lambda_i$ are converged, i.e., $\frac{1}{P}\sum_{i=1}^P \frac{||\lambda_i(t+dt)-\lambda_i(t)||}{ ||\lambda_i(t)||} < 10^{-11}$ where $P$ is the total number of  diagonal matrices.  We can decrease the time step to $dt=0.001$ even smaller  for  further optimization. In order to speed up the convergence, the large-$D$ states are always initialized by a converged PEPS with small $D$ such as $D$=2 rather than random tensors, and what we need to do is just truncating the increased bond dimension to the desired large $D$ when evolution operators act on the $D$=2 state. Once we get  the optimal tensors with SU method, we absorb the environment tensors $\lambda^{1/2}_i$ equally into each local tensor $B_{i}$, used as the initial state for gradient optimization.

\section{Boundary-MPS contraction scheme}
\label{sec:bMPS}

\begin{figure*}[htbp]
 \centering
 \includegraphics[width=6.4in]{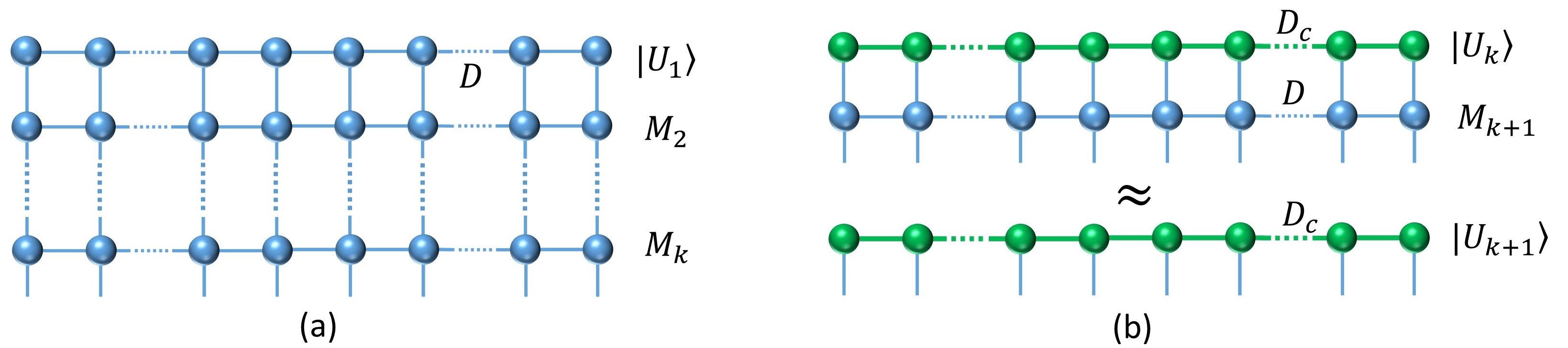}
 \caption{The boundary-MPS scheme for contracting a tensor network. (a) Treat the first row as an MPS $\ket{U_1}$ and the middle rows as MPOs $M_k$, and contract the tensor network row by row from up to down.  (b) During the contraction process, when the MPO $M_{k+1}$ acts on an MPS $\ket{U_k}$, the resulting MPS will be approximated by  $\ket{U_{k+1}}$ with a bond dimension $D_c$. The approximation is performed with a cost $O(L_xD^4D^2_c)$, where $L_x$ is the length of the MPS.}
 \label{fig:boundaryMPS}
 \end{figure*}

In our method, we need to contract a $L_y\times L_x$ single-layer tensor network to get the coefficient $\Psi(S)$ for a given configuration $\ket{S}$.  Here we adopt the boundary-MPS contraction scheme by  contracting the tensor network row by row with treating the first and last row of the tensor network as MPSs and  middle rows as MPOs~\cite{verstraete2008}.  Shown in Fig.~\ref{fig:boundaryMPS}(a), the first row is defined as an MPS $\ket{{U}_1}$ and the $k$th row as an MPO $M_k$. In the contraction process, when the MPO $M_{k+1}$ acts on a given MPS $\ket{U_{k}}$ with a bond dimension $D_c$, we will get a new MPS $\ket{\tilde{U}_{k+1}}=M_{k+1}\ket{U_{k}}$ whose bond dimension is $DD_c$. 

In order to avoid the bond dimension increases exponentially during the contraction process, the resulting MPS  $\ket{\tilde{U}_{k+1}}$ will be approximated by $\ket{U_{k+1}}$ with a smaller bond dimension $D_c$ by minimizing the cost function $f(T_1,T_2,\cdots,T_{L_x})=||M_{k+1}\ket{U_k}-\ket{U_{k+1}}||^2$, where $T_i$ is a rank-3 tensor  related to $\ket{U_{k+1}}$ with dimension $D_c\times D \times D_c$ in the middle sites and $D_c\times D \times 1$ on edges, shown in  Fig.~\ref{fig:boundaryMPS}(b). The solution to  $f(T_1,T_2,\cdots,T_{L_x})$  can be efficiently found  by using an iterative sweep algorithm with cost  O$({L_x}D^4D_c^2)$~\cite{schollw2011}.  Therefore we can contract from up to down and we have $\ket{U_k}=M_kM_{k-1}\cdots M_2\ket{U_1}=M_kM_{k-1}\cdots M_3\ket{{U}_2}=M_k\ket{U_{k-1}}$. Similarly, we can also contract from down to up. Define the last row as $\bra{D_{L_y}}$, then we have $\bra{D_k}=\bra{D_{L_y}}M_{{L_y}-1}\cdots M_{k+1}M_k=\bra{D_{k+1}}M_k$. It is obvious that  $\Psi(S)=\langle D_{k+1}|U_{k}\rangle$ for any $k=1,2,\cdots,L_y-1$.  We can see the cost of contracting the whole tensor network is  O$(ND^4D_c^2)$ where $N=L_y\times L_x$ is the systems size.  

\begin{figure*}[htbp]
 \centering
 \includegraphics[width=6.4in]{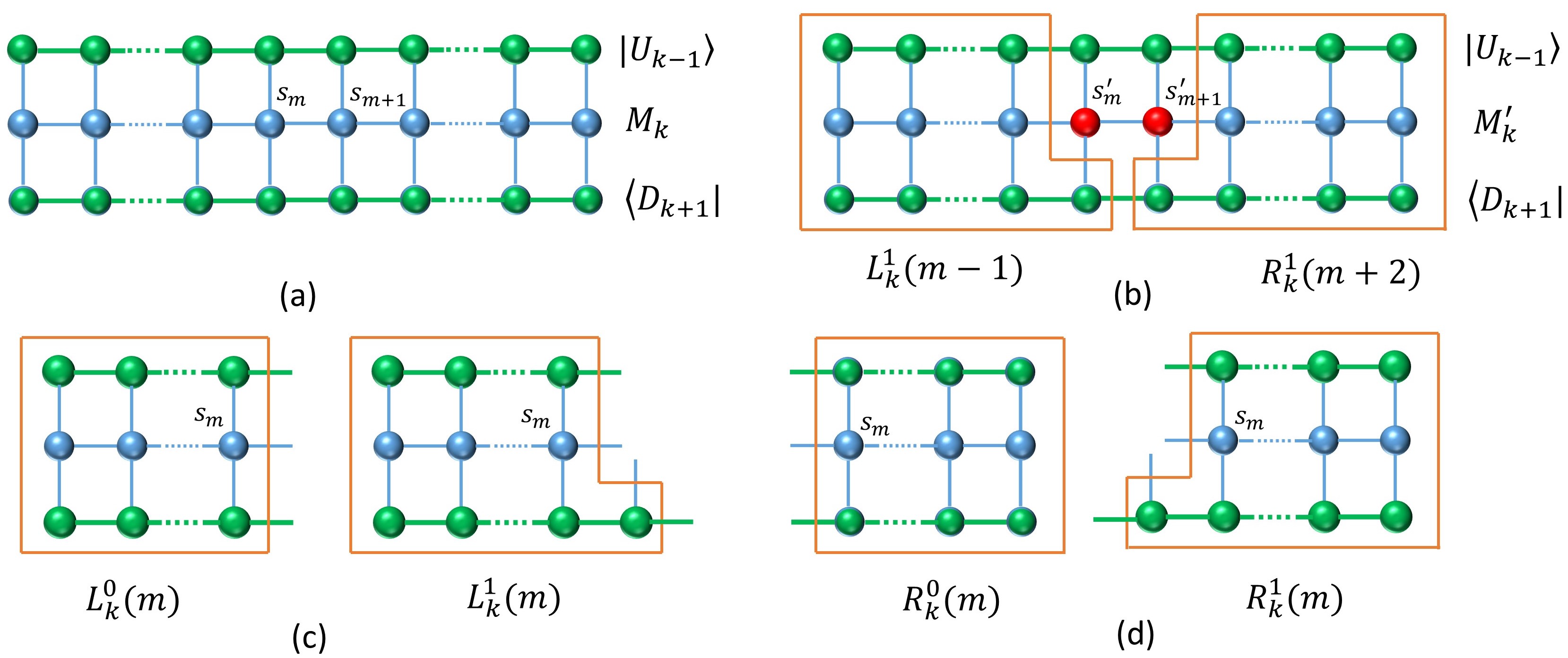}
 \caption{Use auxiliary tensors to efficiently compute energy terms in the $k$-th row. (a) Three-line contraction used to get auxiliary tensors. (b) Replace the $m$-th and $(m+1)$-th tensors of $M_k$ by the other component tensors correspondingly to get another MPO $M^{\prime}_k$.  And use auxiliary tensors $L^{1}_k(m-1)$ and $R^1_{k}(m+1)$ to get the coefficient $\Psi(S^{\prime})$ for computing the horizontal bond energy term between the site $m$ and $m+1$. (c) Contract from the first column to the $m$-th column to get the left auxiliary tensors $L^0_k(m)$, and contract $L^0_k(m)$ with the $(m+1)$-th tensor in $\bra{D_{k+1}}$ to get another auxiliary tensor $L^1_k(m)$. (d)  Contract from the last column to the $m$-th column to get the right auxiliary tensors $R^0_k(m)$, and contract $R^0_k(m)$ with the $(m-1)$-th tensor in $\bra{D_{k+1}}$ to get another auxiliary tensor $R^1_k(m)$.}
 \label{fig:contraction}
 \end{figure*}

\section{Calculations of physical quantities}
\label{sec:energy}

Now we turn to how to calculate the energy for a given state. In the MC sampling, we need to compute the local energy 
 $E_{\rm loc}(S)=\sum_{S^{\prime}} \frac{\Psi(S^{\prime})}{\Psi(S)} \langle S^{\prime}|H|S\rangle $ for a given configuration $|S\rangle=|s_1s_2 \cdots s_N\rangle$. Since the  Hamiltonian is comprised of a series of  NN and NNN  two-body interaction terms, i.e.,  $H=\sum_{\{ij\}}H_{ij}$ and $H_{ij}=\sum_{s_i's_j'}\sum_{s_is_j}(H_{ij})_{s_i's_j',s_is_j}\ket{s_i's_j'}\bra{s_is_j}$,  we have
\begin{equation}
E_{\rm loc}(S)=\sum_{\{ij\}}\sum_{s_i's_j'}\frac{\Psi(S^{\prime})}{\Psi(S)}(H_{ij})_{s_i's_j',s_is_j}, 
\label{eq:Elocal}
\end{equation}
where $\Psi(S^{\prime})$ is the coefficient of the configuration $|S^{\prime}\rangle=|s_1s_2 \cdots s_{i-1}s_i's_{i+1}\cdots s_{j-1}s_j's_{j+1}\cdots s_N\rangle$, and ``$\{\}$'' denotes all NN and NNN spin pairs. The Hamiltonian elements $H_{ij}$ are easily obtained and only nonzero matrix elements  contribute to $E_{\rm loc}(S)$. The main problem for calculating the energy is owing to calculations of  $\Psi(S^{\prime})$ for different spin pairs.  

    It is notable there are at most 1 spin pair in each  configurations $\ket{S^{\prime}}$ different from the configuration $\ket{S}$. This leads to an efficient evaluation for all different  $\Psi(S^{\prime})$ by storing some auxiliary tensors. Taking  the calculation of NN horizontal bond energy terms in the $k$-th row as an example, which contains  ($L_x-1$) NN spin pairs.  We note $\Psi(S)=\langle D_{k+1}|M_k|U_{k-1}\rangle$ and $\Psi(S^{\prime})=\langle D_{k+1}|M^{\prime}_k|U_{k-1}\rangle$,  shown in Fig.~\ref{fig:contraction}(a) and (b). There are only at most two different  tensors between the two MPOs $M^{\prime}_k$ and  $M_k$, denoted by red balls in Fig.~\ref{fig:contraction} (b). 
    
    To get auxiliary tensors,  we contract from the first column to the $m$-th column to get a tensor $L^0_k(m)$, then take $L^0_k(m)$ to contract with the $(m+1)$-th tensor in $\bra{D_{k+1}}$ to get $L^1_k(m)$, shown in Fig.~\ref{fig:contraction}(c). Similarly, we contract from the last column to the $m$-th column and  get a tensor $R^0_k(m)$, then use $R^0_k(m)$ to contract with the $(m-1)$-th tensor in $\bra{D_{k+1}}$ to get $R^1_k(m)$, shown in Fig.~\ref{fig:contraction}(d).  It is easy to verify that $\Psi(S)={\rm Tr}[L^0_k(m)R^0_k(m+1)]$ for $m=1,2,\cdots, L_x-1$.  With the help of auxiliary tensors including $L^0_k(m)$, $L^1_k(m)$, $R^0_k(m)$ and $R^1_k(m)$,  a series of $\Psi(S^{\prime})$  respectively corresponding to the $(L_x-1)$ NN spin pairs can be efficiently computed from left to right scaling as $O(L_xD^4D^2_c)$ , shown in Fig.~\ref{fig:contraction}(b), then  NN horizontal bond energy terms of $E_{\rm loc}(S)$ in the $k$-th row for the given configuration $\ket{S}$ are directly obtained.  At the same time, using the above auxiliary tensors, we can easily get energy gradients  with a leading cost  $O(ND^4D^2_c)$, the same with that of all horizontal bond energy terms. Other energy terms including vertical bond and NNN  terms can be similarly computed.

\section{Convergence of  cutoff $D_c$} 
%\textbf{(Wenyuan, I think this section contains too much details and it might be better still in Appendix)}

%Here we consider the effects caused by cutoff $D_c$. Our method involves two steps when evaluating observables such as energies: generating configurations and computing the local energy $E_{\rm loc}(S)$. The costs of both are dominated by contracting a single-layer tensor network to get $\Psi{(S)}$ scaling as $O(D^4D^2_c)$, and $D_c=3D$ always works very well in practical calculations  for a spin-1/2  $J_1-J_2$ model. We note when generating cofigurations, a not very large $D_c$ can be good enough to produce correct probability distributions, which can further significantly speed up the MC sampling. As we know, the ratio $R=\frac{|\Psi(S_b)|^2}{|\Psi(S_a)|^2}$ in Eq.~(\ref{eq:prob}) is merely used to compare with a random number, and we find the extremely accurate value is not very necessary. This exhibits the intrinsic advantage of MC sampling. Generally, we always set $D_{c1}=2D$ to generate configurations and $D_{c2}=3D$ to compute $E_{\rm loc}(S)$  for both unfrustrated and frustrated cases. The convergence of $D_{c1}$ and $D_{c2}$ is discussed in the following.

\begin{figure}
 \centering
 \includegraphics[width=3.2in]{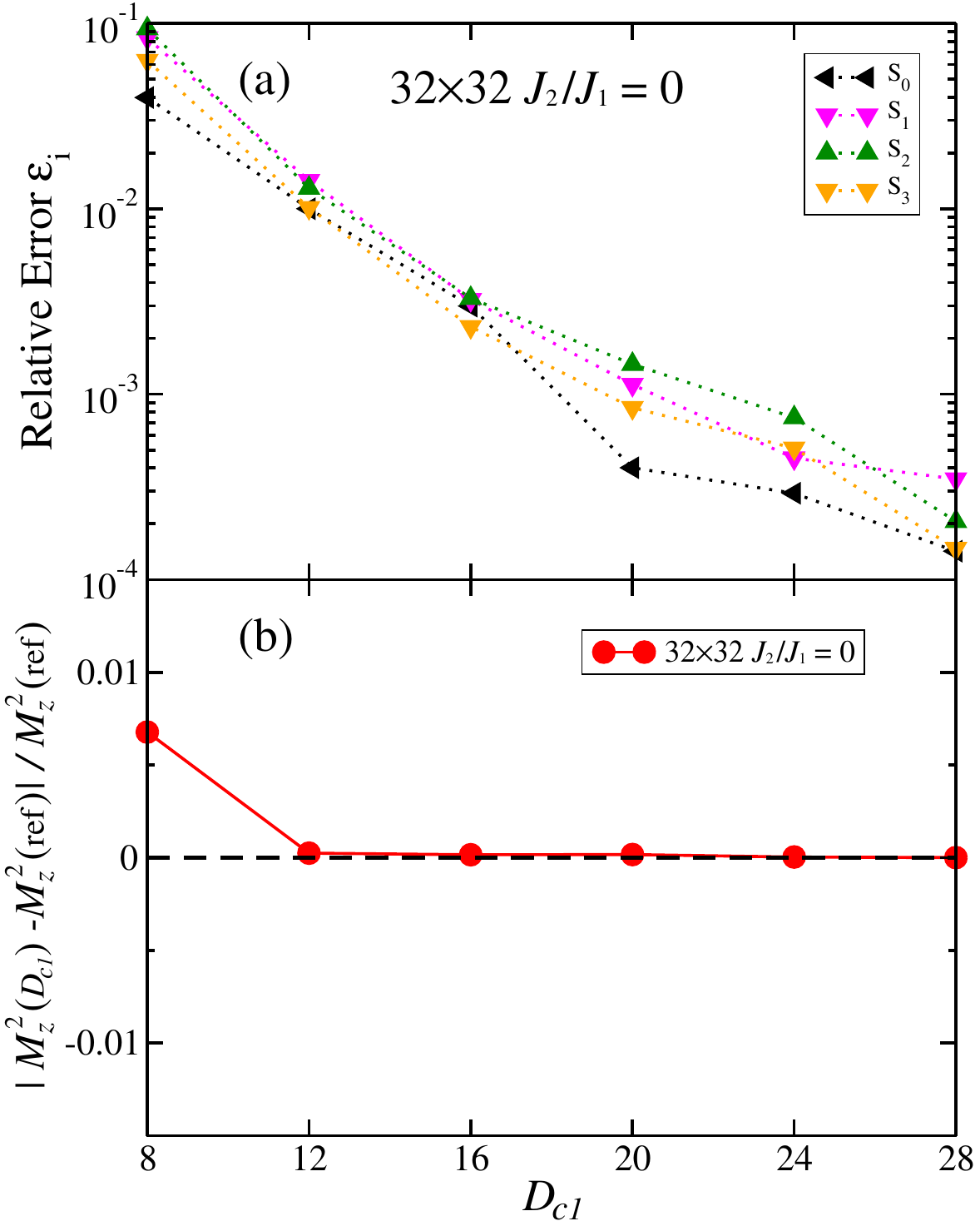}
 \caption{Convergence of the cutoff $D_{c1}$ used for generating configurations on $32\times 32$ Heisenberg model with $D=8$. (a) The relatives error $\varepsilon_i$ of coefficients $\Psi(S_i)$  versus the cutoff $D_{c1}$ for given spin configurations $\ket{S_i}$. Here the configurations  $\ket{S_i}$ are generated randomly and the value $\Psi(S_i)$ of $D_{c1}$=200 is used as the reference.  (b) The relative error of z-component spin orders $M^2_z$ versus the  cutoff $D_{c1}$.  $M^2_z({\rm ref})$ denotes the results of $D_{c1}=32$.  $M^2_z$ is only determined by configurations, therefor measures the correctness of probability distributions of configurations generated with $D_{c1}$.} 
 \label{fig:contractError32}
 \end{figure}

\begin{figure}
 \centering
 \includegraphics[width=3.2in]{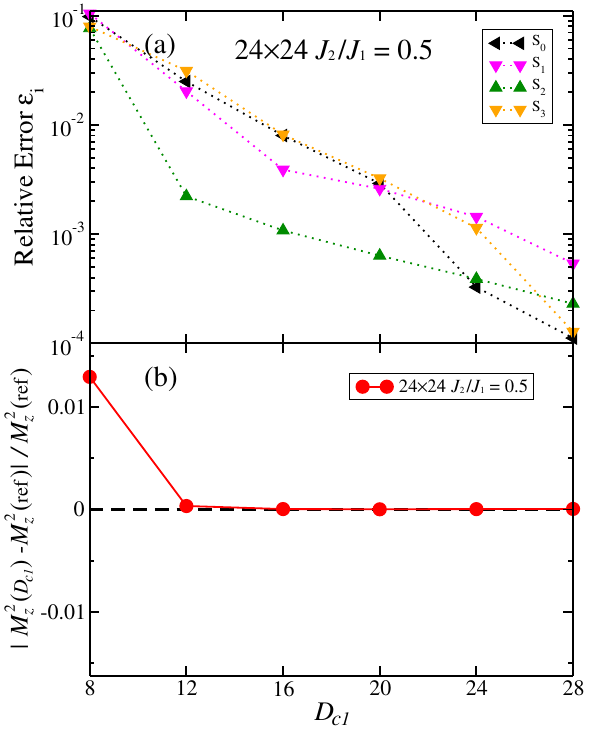}
 \caption{Convergence of the cutoff $D_{c1}$ used for generating configurations on $24\times 24$ at $J_2/J_1$=0.5 with $D=8$. (a) The relatives error $\varepsilon_i$ of coefficients $\Psi(S_i)$  versus the cutoff $D_{c1}$ for given spin configurations $\ket{S_i}$. Here the configurations  $\ket{S_i}$ are generated randomly and the value $\Psi(S_i)$ of $D_{c1}$=200 is used as the reference.  (b) The relative error of z-component spin orders $M^2_z$ versus the  cutoff $D_{c1}$.  $M^2_z({\rm ref})$ denotes the results of $D_{c1}=32$.} 
 \label{fig:contractError24}
 \end{figure}

Here we consider the effects caused by cutoff $D_c$.  When generating configurations, the Metropolis' probability $P_{\rm s}$ is used to compare with a random number $r\in[0,1)$, indicating that it is not necessary to compute the ratio  $R=\frac{|\Psi(S_b)|^2}{|\Psi(S_a)|^2}$ in a high precision. That means a relative small $D_{c1}$ which controls the precision of $R$  is good enough to generate configurations. 
 In order to analyse the influence of $D_{c1}$ on the probability distribution, we define the relative error of $\Psi(S_i)$ for a given configuration $\ket{S_i}$
 \begin{equation}
  \varepsilon_i =\frac{|\Psi(S_i)_{D_{c1}}-\Psi(S_i)_{\rm ex}|}{|\Psi(S_i)_{\rm ex}|} ~,
 \end{equation}
 where $\Psi(S_i)_{D_{c1}}$ is the value using $D_{c1}$ and  $\Psi(S_i)_{\rm ex}$  denotes the exact value for  $\ket{S_i}$.  According to Eq.(\ref{eq:prob}), we can estimate the  error relative to the exact ratio $R_{\rm ex}$ as 
 \begin{equation}
\Big|\frac{\Delta R}{R_{\rm ex}}\Big|=2(\varepsilon_a+\varepsilon_b)\approx 4\varepsilon_0 ~~, 
\end{equation}
where $\varepsilon_a$($\varepsilon_b$) is the relative error of  $\Psi(S_b)$ ($\Psi(S_b)$), and we assume $\varepsilon_a \approx \varepsilon_b \approx \varepsilon_0$.  Then we have  
  \begin{equation}
 P_{\rm s}={\rm min}[1,(1\pm4\varepsilon_0)\cdot R_{\rm ex}] ~~.
\end{equation}
Fig.~\ref{fig:contractError32}(a) depicts the relative error $\varepsilon_i$ of $\Psi(S_i)$ for a $32\times 32$ Heisenberg model, where the configurations  $\ket{S_i}$ are generated randomly. We can see the relative errors for different $\ket{S_i}$ are roughly in the same order. For $D_{c1}$=8, the relative error of $\Psi(S_i)$  is about 10\%, correspondingly the relative error of $R$ is about 40\%. When $D_{c1}$ gets larger such as $D_{c1}$=12, the error $\Psi(S_i)$ is reduced to 1\% with the relative error of $R$ being 4\%, which makes it possible to produce the correct probability distribution. The relative error of $\varepsilon_i$ of $\Psi(S_i)$ for $24\times24$ at $J_2/J_1$=0.5 shows a similar behaviour, shown in Fig.~\ref{fig:contractError24}(a). We note the z-component antiferromagnetic spin order $M^2_z=\frac{1}{N^2}\sum_{ij}(-1)^{i+j}\langle S^z_iS^z_j\rangle $ can be used to detect whether the probability distribution is correctly generated, because $M^2_z$  only dependens on spin configurations. Configurations are generated by different $D_{c1}$ and the corresponding relative error of $M^z_2$ is computed and depicted as a function of $D_{c1}$ in Fig.~\ref{fig:contractError32}(b) and Fig.~\ref{fig:contractError24}(b). We can see the relative error of $M^2_z$ is almost zero for $D_{c1}\geq 16$, indicating $D_{c1}$=16 is enough to produce a correct probability distribution for $32\times 32$ at $J_2/J_1$=0 and $24\times 24$ at $J_2/J_1$=0.5. Therefore, without loss of generality, we can use  $D_{c1}=2D$ to generate configurations in practical calculations.  

\begin{figure}
 \centering
 \includegraphics[width=3.2in]{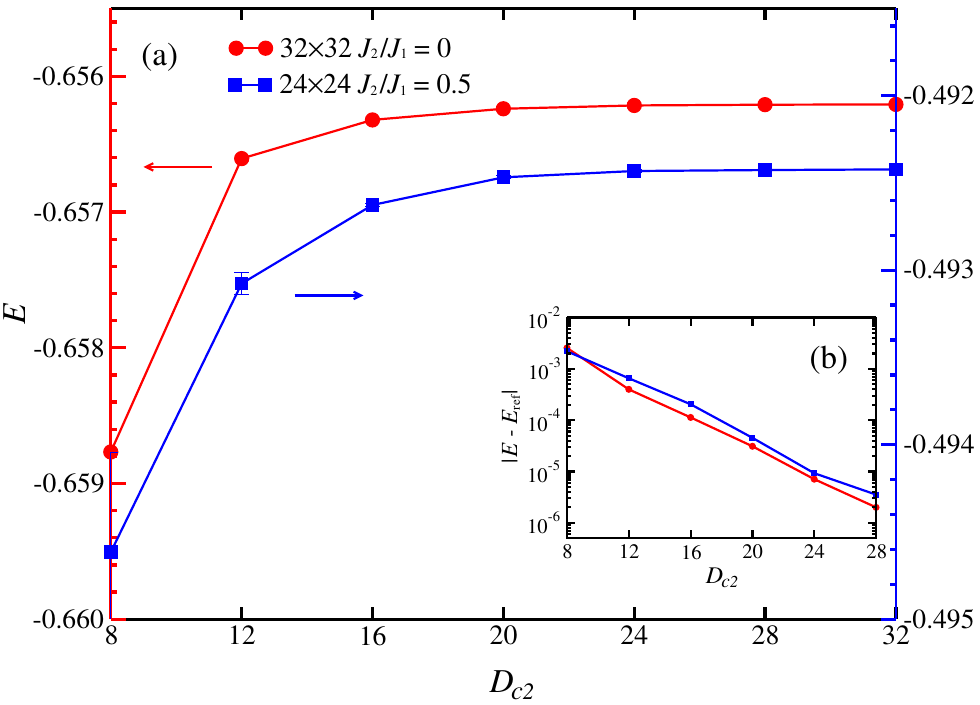}
 \caption{(a) The convergence of energy versus the cutoff $D_{c2}$ on a $32\times 32$ Heisenberg model and  $24\times 24$ $J_2/J_1$=0.5 model.  (b) The absolute error of energy versus $D_{c2}$. $E_{\rm ref}$ denotes the energy using $D_{c2}=32$.}
 \label{fig:Dcutconverge}
 \end{figure}

 We know the precision of physical observables including  energy and spin correlations depends on the value of  $\frac{\Psi(S^{\prime})}{\Psi(S)}$ according to Eq.(\ref{eq:Elocal}). That means different from generating configurations  where a relatively small cutoff $D_{c1}$ can work well, a  large cutoff $D_{c2}$ is needed to compute observables if a high precision is required. Taking the energy calculation as an example, Fig.~\ref{fig:Dcutconverge}(a) shows the energy convergence with respect to $D_{c2}$ on the $32\times 32$ Heisenberg model and $24\times 24$ at $J_2/J_1$=0.5. We can see in both cases when increasing $D_{c2}$ from 8 to 20 the energy converges very fast, while changes very little for $D_{c2}\geq 20$. Compared with the result of $D_{c2}=32$, the energy errors of $D_{c2}=24$ is as small as $7.1\times 10^{-6}$ for $32\times 32$ and $9.2\times 10^{-6}$ for $24\times 24$ lattice, which are the same order of MC sampling errros, show in Fig.~\ref{fig:Dcutconverge}(b). That means $D_{c2}=3D$ works quite well to compute physical quantities in the range of allowed MC sampling errors. One can also use larger $D_{c1}$ and $D_{c2}$, but we do not find improvements in our calculations.

\section{MC sampling convergence}
\begin{figure}
 \centering
 \includegraphics[width=3.2in]{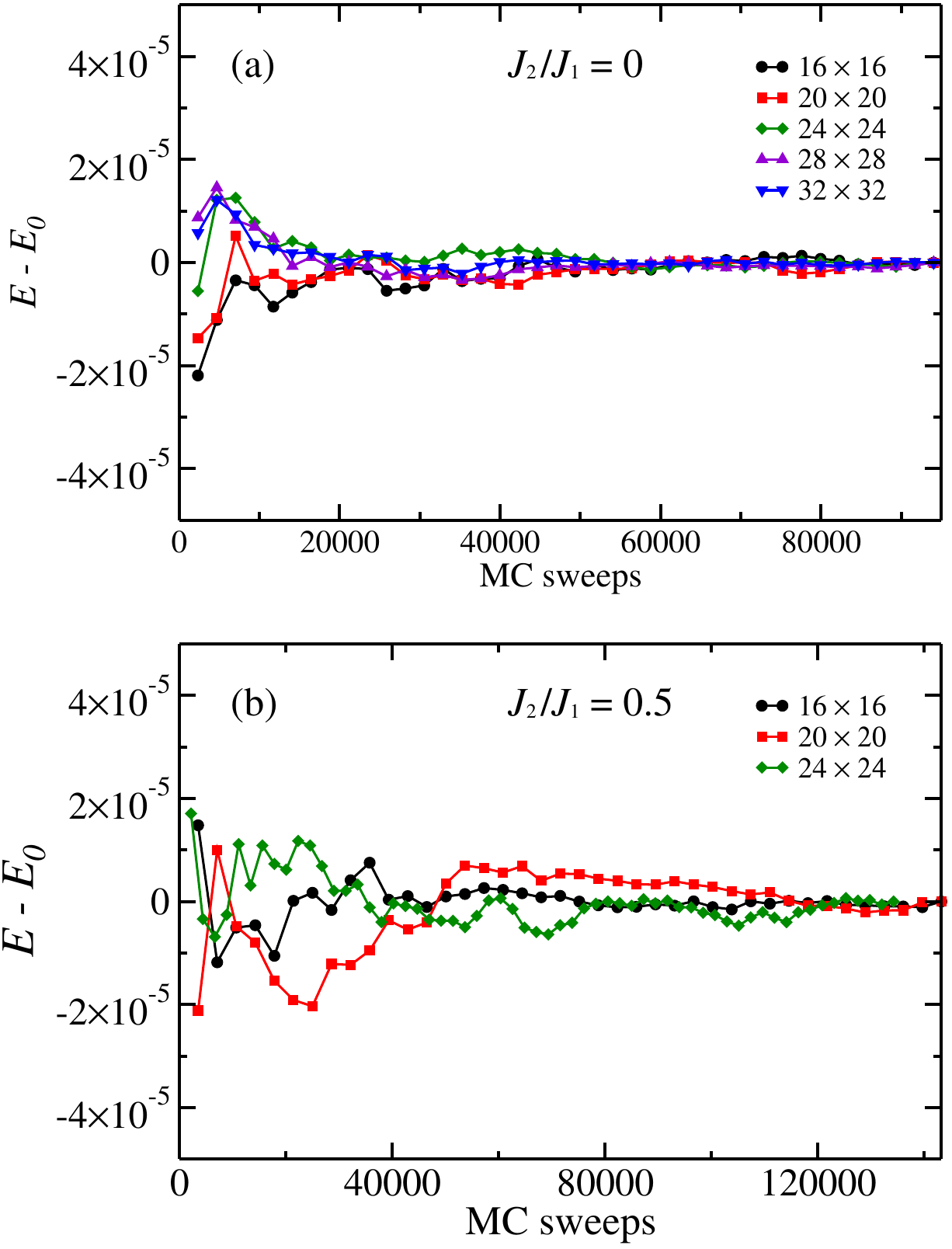}
 \caption{The  energy convergence versus MC sweeps for different systems $L\times L$ (a) on Heisenberg model and (b) at $J_2/J_1$=0.5. Here $E_0$ denotes the final energy with maximal MC sweeps. } 
 \label{fig:MCconverge}
 \end{figure}

We compare the energy convergence with respect to MC sampling on different systems in the scheme of sequentially visiting spin pairs using $D=8$. Here we use $D_{c1}=2D$ to generate configurations and  $D_{c2}=3D$ to compute energy $E_{\rm loc}(S)$ for a given configuration $\ket{S}$.  Fig.~\ref{fig:MCconverge} depicts how the energy per site $E$ changes versus the number of MC sweeps for unfrustrated and frustrated models. To directly compare the energy convergence of different sizes, we subtract their corresponding final energy $E_0$ which is just the energy with maximal MC sweeps. At $J_2/J_1$=0, 20000 MC sweeps can converge the energy 
within errors $1.0\times 10^{-5}$, and at $J_2/J_1$=0.5 about 50000 MC sweeps are needed to get the same precision. We note in either unfrustrated or frustrated case, all different systems show  the similar convergence behavior, namely, the convergence with respect to MC sweeps is almost size independent.  That means our method can be directly applied to larger systems by using more computational resources.

%\bibliography{PEPS_OBC_revised}

\end{document}